\documentclass[apj]{emulateapj}

\usepackage{graphicx}
\usepackage{times}
\usepackage{amsmath}
\usepackage{color}
\usepackage{graphics}
\usepackage{subfigure}
\usepackage{txfonts}
\usepackage{natbib}
\usepackage{wasysym}

\bibpunct{(}{)}{;}{a}{}{,}

\shorttitle{Critical decay index at the onset of solar eruptions}
\shortauthors{F.P. Zuccarello et al.}

\begin{document}


         
\title{{Critical decay index at the onset of solar eruptions}}

\author{F.P. Zuccarello}
\email{Francesco.Zuccarello@obspm.fr}

\author{G. Aulanier}

\email{Guillaume.Aulanier@obspm.fr}
\author{S.A. Gilchrist}

\email{Stuart.Gilchrist@obspm.fr}
\affil{LESIA, Observatoire de Paris, PSL Research University, CNRS, Sorbonne Universit\'{e}s, UPMC Univ. Paris 06, Univ. Paris Diderot,
Sorbonne Paris Cit\'{e}, 5 place Jules Janssen, 92195 Meudon, France.}

\begin{abstract}

Magnetic flux ropes are topological structures consisting of twisted magnetic field lines that globally wrap around an axis. 
The torus instability model predicts that a magnetic flux rope of major radius $R$ undergoes an eruption when its axis 
reaches a location where the decay index $-d(\ln B_{ex})/d(\ln R)$ of the ambient magnetic field $B_{ex}$ is larger than a 
critical value. In the current-wire model, the critical value depends on the thickness and time-evolution of the current channel. 
We use magneto-hydrodynamic (MHD) simulations to investigate if the critical value of the decay index at the onset of the eruption 
is affected by the magnetic flux rope's internal current profile and/or by the particular pre-eruptive photospheric dynamics. 
The evolution of an asymmetric, bipolar active region is driven by applying different classes of photospheric motions.
We find that the critical value of the decay index at the onset of the eruption is not significantly affected by either the 
pre-eruptive photospheric evolution of the active region or by the resulting different magnetic flux ropes. 
As in the case of the current-wire model, we find that there is a `critical range' $ [1.3-1.5]$, rather 
than a `critical value' for the onset of the torus instability. This range is in good agreement with the 
predictions of the current-wire model, despite the inclusion of line-tying effects and the occurrence 
of tether-cutting magnetic reconnection.

\end{abstract}


\section{Introduction}

Solar eruptions are one of the most spectacular and violent phenomena 
that occur in the Sun's atmosphere. Together with solar flares they are 
the most impulsive and energetic manifestation of solar activity. 
Energy considerations suggest that eruptions, and the associated coronal-mass 
ejections (CMEs), are magnetically driven \citep{Forbes2006}. 

A significant number of eruptions originate in active regions where the 
magnetic field is significantly sheared. Many of these active regions 
also host filaments, that is, structures consisting of plasma that is 
cooler and denser than its surroundings.


Magnetic flux ropes, i.e., twisted magnetic field lines that globally wrap 
around an axial magnetic field, are topological structures that can host 
and govern the dynamics of filaments \citep{Kup1974,van1989,Pri1989,Dem1989,Aul1998,van2004,Mac2006,Gun2015}.
In addition, they can also explain the observed sigmoidal/highly-sheared structures  
of erupting active regions \citep{Can2010,Jing2010,Gre2011,Sav2012,Gibb2014,Jiang2014}. 
Structures compatible with magnetic flux ropes have also 
been found in solar cavities \citep{Gib2010,Rac2013}.


This evidence suggests that magnetic flux ropes play a fundamental role 
in solar eruptions, and explains why virtually all CME-initiation models 
have a phase in which a magnetic flux rope is present \cite[see reviews by][]{Forbes2010,Chen2011,Aul2014,Fil2015}. 

Based on the observation that many filament eruptions are associated 
with convergence motions towards the polarity-inversion line (PIL) 
of active regions, \cite{van1989} proposed flux cancellation at the PIL 
as a possible mechanism for the formation and eruption of magnetic flux ropes. 
This eruption scenario has been further investigated in 2D by \cite{For1991} 
and \cite{Ise1993} by using an ideal magneto-hydrodynamic (MHD) description 
of a magnetic flux rope embedded in the field generated by a sub-photospheric 
line dipole. In particular, these authors studied the evolution of the 
system in the equilibrium manifold when convergence flows towards the PIL 
were applied. In response to this photospheric driver, the magnetic flux 
rope follows a series of nearby equilibria that are located at larger 
and larger heights. This evolution continues until a critical point is 
reached. At this point, no nearby equilibrium is accessible and the system 
experiences a catastrophic \textit{loss of equilibrium}: the magnetic 
flux rope suddenly jumps to a new equilibrium at a significantly larger 
height, and eventually experiences a full eruption if magnetic reconnection 
is allowed \citep{Lin2000}. 

Another ideal-MHD mechanism for the initiation of CMEs is the 
\textit{torus instability} \citep{Bat1978,Kliem2006}. In this model, a 
current ring of major radius $R$ is embedded in an external magnetic 
field. Due to the curvature of the current channel, the ring experiences 
a radial `hoop force', which is directed outwards and decreases in 
magnitude if the ring expands. If the inwardly directed 
Lorentz force due to the external field decreases faster with $R$ than 
the hoop force, the system becomes unstable.

Assuming an external field $B_{ex} \propto R^{-n}$, the decay index $n$ is defined as
\begin{equation}
n=-\frac{d \ln B_{ex}}{d \ln R}
\label{Eq1}
\end{equation}
\cite{Bat1978} and \cite{Kliem2006} showed that this instability occurs 
when $n \geqslant n_{crit} =1.5$. In other words, if the current ring 
has a major radius, R, such that the decay index of the external field, 
$n$, is significantly smaller than $n_{crit}$, the system is in a stable 
equilibrium where the inwardly-directed magnetic tension of the external field 
balances the outwardly-directed magnetic pressure of the current channel. 
However, if $n \geqslant  n_{crit}$, then this equilibrium is unstable and 
any displacement of the current channel due to some perturbation will 
initiate an outward motion of the current ring.

By using a current-wire approach, \cite{Dem2010} have shown that
the torus instability is equivalent to a loss of equilibrium.
In fact, the torus instability is the instability that occurs at 
the critical point of the equilibrium manifold of the loss of equilibrium model \citep{Kliem2014}.

The exact critical value of the decay index, $n_{crit}$, at which the loss 
of equilibrium occurs depends on the morphology of the current wire. 
If thin current distributions are considered, then
$n_{crit} = 1$ for an infinitely long wire, and $n_{crit} = 1.5$  for 
a perfectly circular current ring. If relatively thick current distributions 
are considered, then the difference between $n_{crit}$ for a 
straight wire and a circular ring is smaller. 
The current distributions found in MHD calculations are typically thick: 
the radius of the cross section is significant compared to the length
of the current channel. For relatively thick current distributions
the critical decay index lies in the range 
$n_{crit}=1.1-1.3$ or $n_{crit}=1.2-1.5$ depending on whether or not the
current wire expands during the perturbation \citep{Dem2010}. Independently 
of the exact topology of the external field, \cite{Kliem2014} have found 
that for a T\&D \citep{Tit1999} flux rope the instability occurs when the 
critical decay index is $n_{crit} \simeq 1.4$. \cite{Olm2010} analytically 
investigated  the stability properties of a line-tied partial torus and 
found that the critical decay index for the onset of the instability is 
in the range of $n_{crit} \approx 0.5-2$ depending on the fraction 
of the torus ---from half to a full torus---  that is above the photosphere. 

The aforementioned results are derived using wire models, that is, 
the equilibrium properties are determined by solving the momentum 
equation in terms of the current distribution. In recent years, many 
numerical experiments have been performed to validate the torus 
instability model using the full set of  MHD equations. \cite{Tor2005,Tor2007} 
performed numerical MHD simulations of a line-tied T\&D flux rope embedded 
in different external magnetic field configurations and found that at the 
moment of the eruption the decay index at the apex of the flux rope axis 
is $n_{crit} \simeq 1.5$. A similar value was found in a simulation where 
the magnetic flux rope is dynamically formed through magnetic reconnection 
at bald-patches and at hyperbolic flux tubes \citep{Aul2010}. However, flux 
emergence simulations seem to suggest a higher value of the critical decay 
index. For example, \cite{Fan2007} and  \cite{Fan2010} have found that the 
critical decay index at the onset of the eruption lies in the range 
$n_{crit}=1.75-1.9$, while \cite{An2013} found values for $n_{crit}$ 
that are well above 2. Recently, MHD relaxations of NLFF equilibria of 
solar active regions suggest values of the critical decay index in the 
range $n_{crit}=1.5-1.75$ \citep{Kliem2013,Ama2014,Ino2015}. 

It is clear that different studies give rise to significantly different 
critical values of the decay index for the onset of the instability. 
Therefore, it is natural to ask why so many different values exist.

Many of the basic concepts developed for the torus instability 
(such as the decay index) were developed in the current-wire framework. 
Applying these concepts directly to MHD calculations is difficult 
due to the differences between the two approaches.
Firstly, there is the problem of identifying \textit{where} to evaluate 
the decay index, $n$, to compare to the critical value $n_{crit}$. There
is no ambiguity in the current-wire models -- the decay index is always 
computed at the apex of the infinitesimal current wire. However, 
an equivalent structure does not generally exist in an MHD simulation, 
so it is unclear where to evaluate Eq.~(\ref{Eq1}). A natural choice is 
to follow the idealized case and compute $n$ at the axis of the flux rope, 
but generally the axis is not well-defined unless the rope is symmetric.
Secondly, it is difficult to determine which $n_{crit}$ should be used.
In the current-wire formulation, different values of $n_{crit}$ are 
derived for different prescriptions of the current profile in the wire. Generally, 
the current distribution of an MHD calculation will not match any of 
the idealized configurations. Finally,
the current-wire approach often does not include the effect of
line-tying, which also affects $n_{crit}$. Given these problems, one might
expect the two approaches to predict significantly different values
for the onset of the instability.

However, and even more importantly, different MHD simulations result in different values of the critical decay index, raising further questions such as what is the role of different line-tied photospheric drivers? Does the morphology of the magnetic flux rope also influence the critical value of the decay index in the MHD treatment? Does the current distribution within the flux ropes, that differs in different simulations, affect the critical value of the decay index?  Can the different critical values of the decay index be due to the identification of the axis of the magnetic flux rope? 

In order to address these questions, in the present paper, we perform a parametric study aimed to determine the value of the critical decay index at the onset of the loss of equilibrium, when different classes of photospheric motions --- that resemble the ones typically observed in active regions --- are applied. Starting from an asymmetric, bipolar active region as in \cite{Aul2010}, we apply four different classes of motions, namely convergence toward the active region's PIL, asymmetric stretching, and peripheral and global dispersal of the active region. We describe how the corona responds to the different drivers and how these drivers affect the height and only marginally the critical value $n_{crit}$, of the onset of the torus instability.  

The plan of our paper is as follows. In the following section we introduce our numerical model, the initial condition and the implementation of the   boundary conditions. The topological and energy evolution of the system in response to these flows is presented in Section~\ref{Evolution}. Section~\ref{TI} describes the analysis that we performed in order to determine the critical value of the decay index at the onset of the eruptions. Finally, in Section~\ref{Conclusion} we discuss our findings and conclude.

\section{Model Setup}

The dynamics of the formation and evolution of magnetic flux ropes is modelled by using a new hybrid MPI/OpenMP parallel version of the Observationally-driven High-order Magnetohydrodynamics code \citep[OHM,][]{Aul2005,Aul2010}. The OHM-MPI code solves the following  zero-$\beta$ (pressureless), time-dependent MHD equations in Cartesian coordinates:

\begin{equation}
\frac{\partial \rho}{\partial t} = - \nabla \cdot (\rho \mathbf{u}) + \xi \Delta (\rho -\rho_0),
\end{equation}
\begin{equation}
\rho \frac{\partial \mathbf{u}}{\partial t} = - \rho~ (\mathbf{u} \cdot \nabla) \mathbf{u} + \mathbf{J} \times \mathbf{B} + \rho \nu' \mathcal{D}\mathbf{u},
\end{equation}
\begin{equation}
\frac{\partial \mathbf{B}}{\partial t} = \nabla \times (\mathbf{u} \times \mathbf{B}) + \eta \Delta \mathbf{B}  ,
\end{equation}
\begin{equation}
\nabla \times \mathbf{B} = \mu\mathbf{J}  ,
\end{equation}
where $\rho$ is the mass density ($\rho_0$ is its initial value at $t = 0$), $\mathbf{u}$ is the plasma velocity, $\mathbf{B}$ is the magnetic field, $\mathbf{J}$ is the electric current-density, $\eta$ is the magnetic resistivity and $\mu$ is the magnetic permeability of the vacuum. $\xi\Delta(\rho-\rho_0)$ and $\nu' \mathcal{D}$ are artificial diffusion operators for the density and the velocity, whose presence is necessary to ensure the numerical stability of the code \citep[see][for details]{Aul2005}. Furthermore, in the continuity equation an extra `Newton's term' is added for $z \in [0,0.7]$ in order to avoid sharp density variations close to the bottom boundary. 

The calculation is parallelized using a combination of the OpenMP 
\citep{Chandra2001} and the Message-Passing Interface (MPI) standards \citep{Gropp1999}. 
For OHM-MPI, OpenMP is used to distribute work among cores on individual 
nodes, while MPI is used to pass data between nodes. 
The computational volume is sliced horizontally (i.e.\ in $x-y$ plane) to produce a vertical stack of sub-volumes. 
Each node is assigned the task of computing the solution in 
a particular sub-volume (which is itself divided among the cores 
on the  particular node). At each time step, synchronization and data transfer between nodes 
is necessary to compute derivatives across sub-volume boundaries. 
The parallelization achieves a significant 
speed-up for a moderate number of cores ($\sim$100), however for 
large numbers of cores, the calculation is limited by the finite memory bandwidth 
in the shared-memory environment, and the communication times in the 
distributed memory environment. 

The three-dimensional MHD equations are solved on a non-uniform mesh that covers the physical domain $ [-10,10] \times[-10,10] \times [0,30]$ using $251\times251\times231$ grid points with a grid resolution that varies in the range $[6 \times 10^{-3} , 0.32] \times [6 \times 10^{-3} , 0.32] \times [6 \times 10^{-3}, 0.6]$, and with the smallest cell centered at $x=y=z=0$. Similarly to \cite{Aul2010}, the MHD equations are solved in their dimensionless form. 


\subsection{Initial Condition}
As initial condition for the magnetic field we consider the current-free (potential) field generated by two unbalanced monopoles placed at a distance $L=2$ from each other and located at different depths below the photosphere ($z=0$). The mathematical form of this field is \citep{Aul2010}:
\begin{align}
\nonumber &B_x(t=0)=\sum \nolimits_{i=1}^2 C_i~(x-x_i)~r_{i}^{-3}, \\
 &B_y(t=0)=\sum \nolimits_{i=1}^2 C_i~(y-y_i)~r_{i}^{-3},          \\
\nonumber &B_z(t=0)=\sum \nolimits_{i=1}^2 C_i~(z-z_i)~r_{i}^{-3},  \\
\nonumber r_i&=\sqrt{(x-x_i)^2+(y-y_i)^2+(z-z_i)^2}
\end{align}
where ($x_1 =1.025$;~$y_1 =0.3$;~$z_1 =-0.9$;~$C_1 =15$) and ($x_2 = -0.775$;~$y_2 =-0.3$;~$z_2 = -1.3$;~$C_2 = -14$). This initial condition results in an asymmetric active region with a positive flux excess of about 24\%.

The initial density profile is defined as 
\begin{equation}
\rho(t=0) = B^2(t=0),
\end{equation}
resulting in an uniform Alfv\'{e}n speed ($c_A =1$) everywhere in the computational domain. This ensures that if the boundary motions are sub-Alfv\'{e}nic at the photosphere ($z=0$), then they are sub-Alfv\'{e}nic everywhere, therefore avoiding the generation of steep wave fronts at larger heights. 
The initial velocity field is $\mathbf{u} =0$ in the entire computational domain.

\subsection{Boundary Conditions} \label{boundary}

We impose zero-gradient, `open' boundary conditions for all the MHD variables at all the boundaries except for the boundary at $z=0$, where we impose the so-called `line-tied' boundary conditions, therefore ensuring that the foot points of each magnetic field line do not move unless the motions are explicitly prescribed \citep{Aul2005}. 

In order to  quasi-statically evolve the initial potential magnetic field into a current-carrying magnetic field, at the line-tied boundary we impose the following velocity profile:
\begin{equation}
\mathbf{u}(t) = \gamma(t)~\mathbf{u}_0(t),
\label{twist}
\end{equation} 
\begin{align}
\nonumber &\mathbf{u}_0(t) = u_0^{\text{max}}~\psi_0(t)~ \left[\nabla_{\perp} \psi(t) \right] \times \mathbf{e}_z, \\
&\psi(t) = \exp \left[-\psi_1\left(\frac{B_z(z=0;t)}{B_z^{\text{min}}(z=0;t)}\right)^2\right],
\label{psi}
\end{align}
where $\psi_1 = 3.5$, $u_0^{\text{max}} =0.05c_A$ and $\psi_0(t)$ is computed at every time step to guarantee that the maximum value of $\mathbf{u}_0(t)$ is always equal to $u_0^{\text{max}}$, therefore ensuring the sub-Alfv\'{e}nic character of the driving motions. The function
\begin{equation}
\gamma(t) = \prod \nolimits_{i=0}^1 \left[  \frac{( -1 ) ^ {i}}{2} \tanh \left(2~\frac{t-t_i}{\Delta t} \right) + \frac{1}{2} \right]
\label{gamma}
\end{equation}
is used to ensure that the flows are smoothly increased (respectively, decreased) from zero to their steady value (respectively, from their steady value to zero) within a time interval $2\Delta t$ centred at $t=t_0$ (respectively, $t=t_1$). In what follows, we will only give the time $t_i$ as the time when the flows and/or diffusion are switched-on (switched-off), however implying that this is achieved over a time interval $[t_i-\Delta t, t_i+\Delta t]$. 

We choose $t_0=10 t_A$, $t_1=100 t_A$ and $\Delta t=3t_A$, that is, the photospheric motions are applied until $t=100t_A$ when they decrease to half of their steady value and become infinitesimal after $t=103t_A$. 

By construction these flows are asymmetric vortices centred around the local maxima of $|B_z|$, with the fastest velocity being reached close to the PIL and rapidly decreasing when moving away from it and toward the center of the magnetic field polarities. As a consequence, these flows induce shear close to the PIL, without significantly perturbing the magnetic field  anchored around the center of the magnetic polarities.  

In order to study the effect of different photospheric flows on the formation and stability of magnetic flux ropes, starting from time $t=105t_A$ we impose four different classes of flows resulting in four different simulations runs. More precisely, convergence of the magnetic flux closest to the PIL (`Run C': convergence), asymmetric stretching along one direction only (`Run S': stretching), and peripheral and global dispersal of the magnetic field polarities (`Run D1, D2': dispersal). 

To define these boundary conditions we use Equation~(\ref{twist}) where $\gamma(t)$ is given by Equation~(\ref{gamma}), with $t_0=105t_A$, $\Delta t=3t_A$, and $t_1$ the time when the eruption becomes unavoidable. The time $t_1$ is determined through a series of relaxations runs (see Section \ref{relax}). Finally, $\mathbf{u}_0(t) \equiv [u_x(t),u_y(t)]$ is defined as:
\begin{itemize}
 \item  `Run C': 
\begin{align}
   \nonumber u^{\text{C}}_x(t) &=
    \begin{cases}
      u_0^{\text{max}}~\psi_0(t)~\partial_x \psi(t), & \text{if}\ u_x(t) \cdot B_z(t) \leq 0 \\
      ~~~~~~~~~~~~0, & \text{otherwise}
    \end{cases} 
    \\
    u^{\text{C}}_y(t) &= 0
    \label{mC}
  \end{align}
\item  `Run S':
\begin{align}
   \nonumber u^{\text{S}}_x(t) &= u_0^{\text{max}}~\psi_0(t)~\partial_x \psi(t) ~~~~~~~~~~~~~~~~~~~~~~~~~~~~~~~~~~~~\\
    u^{\text{S}}_y(t) &= 0
\end{align}
\item   `Run D1': 
\begin{align}
   \nonumber u^{\text{D1}}_x(t) &= u_0^{\text{max}}~\psi_0(t)~\partial_x \psi(t) ~~~~~~~~~~~~~~~~~~~~~~~~~~~~~~~~~~~~\\
    u^{\text{D1}}_y(t) &= u_0^{\text{max}}~\psi_0(t)~\partial_y \psi(t)
\end{align}
\end{itemize}
with $\psi(t)$, $\psi_0(t)$, $\psi_1$, $u_0^{\text{max}}$ the same as in Equations~(\ref{psi}) and
\begin{itemize}
\item   `Run D2': 
\begin{align}
   \nonumber u^{\text{D2}}_x(t) &= u_0^{\text{max}}~\psi_0(t)~\partial_x \psi(t) ~~~~~~~~~~~~~~~~~~~~~~~~~~~~~~~~~~~~\\
    u^{\text{D2}}_y(t) &= u_0^{\text{max}}~\psi_0(t)~\partial_y \psi(t)
    \label{mD2}
\end{align}
\end{itemize}
that is, exactly the same as `Run D1',  but with $\psi_1=-0.5$.  

\begin{figure}
\begin{center}
\subfigure{\label{RunC-vfield}
\includegraphics[width=.47\textwidth,viewport=6 4 793 392,clip]{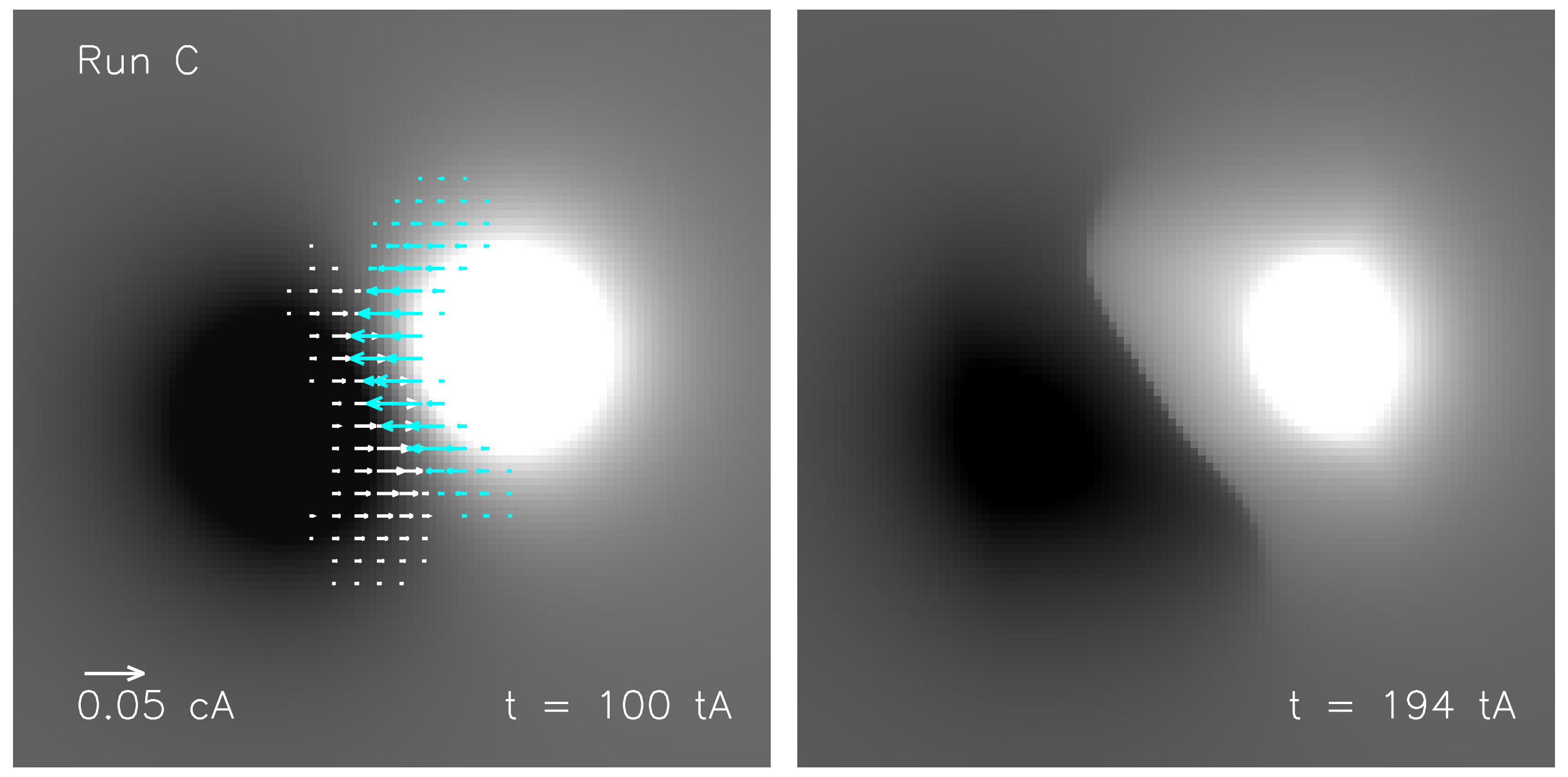}}
\subfigure{\label{RunS-vfield}
\includegraphics[width=.47\textwidth,viewport=6 4 793 392,clip]{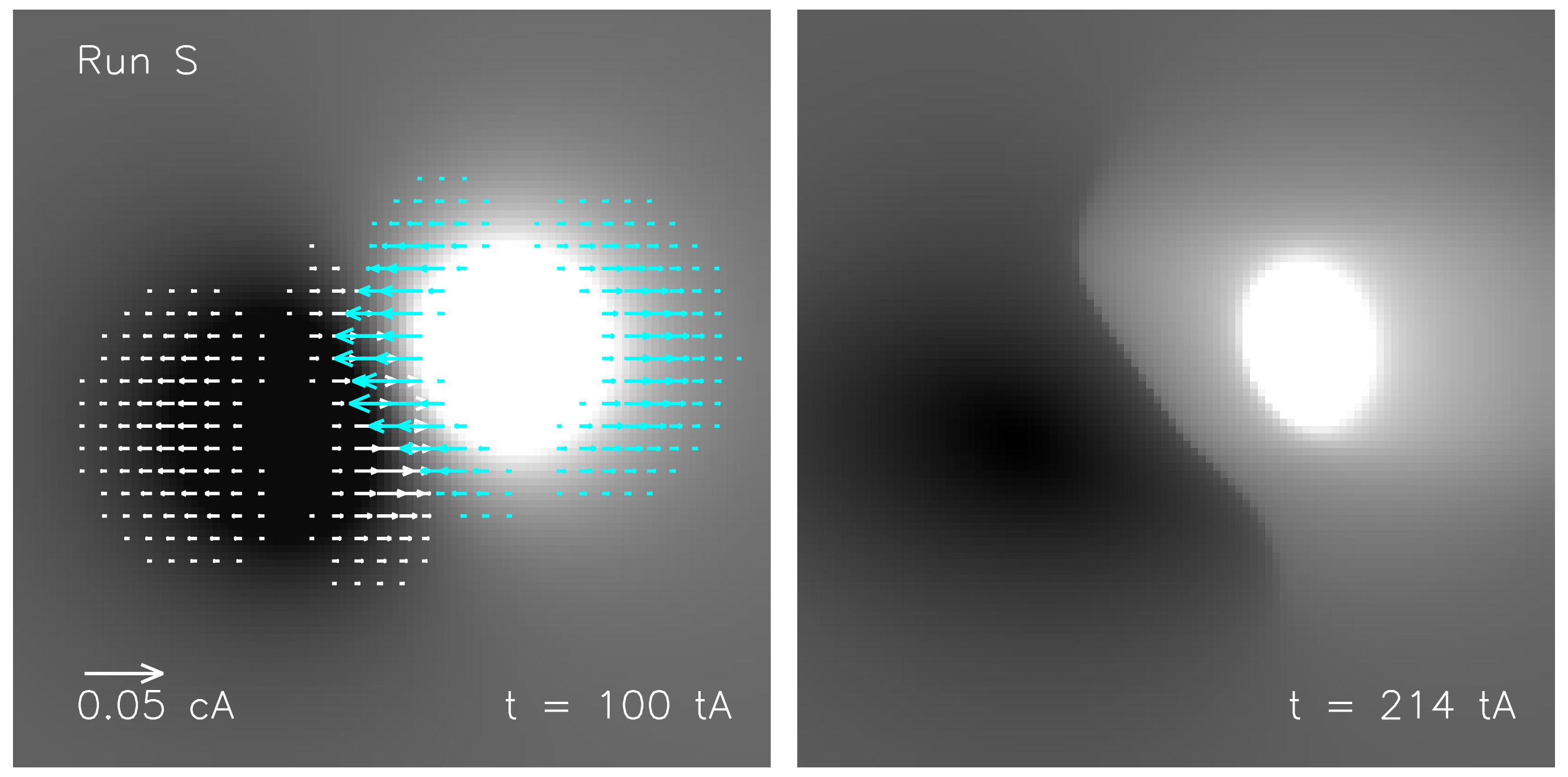}}
\subfigure{\label{RunD1-vfield}
\includegraphics[width=.47\textwidth,viewport=6 4 793 392,clip]{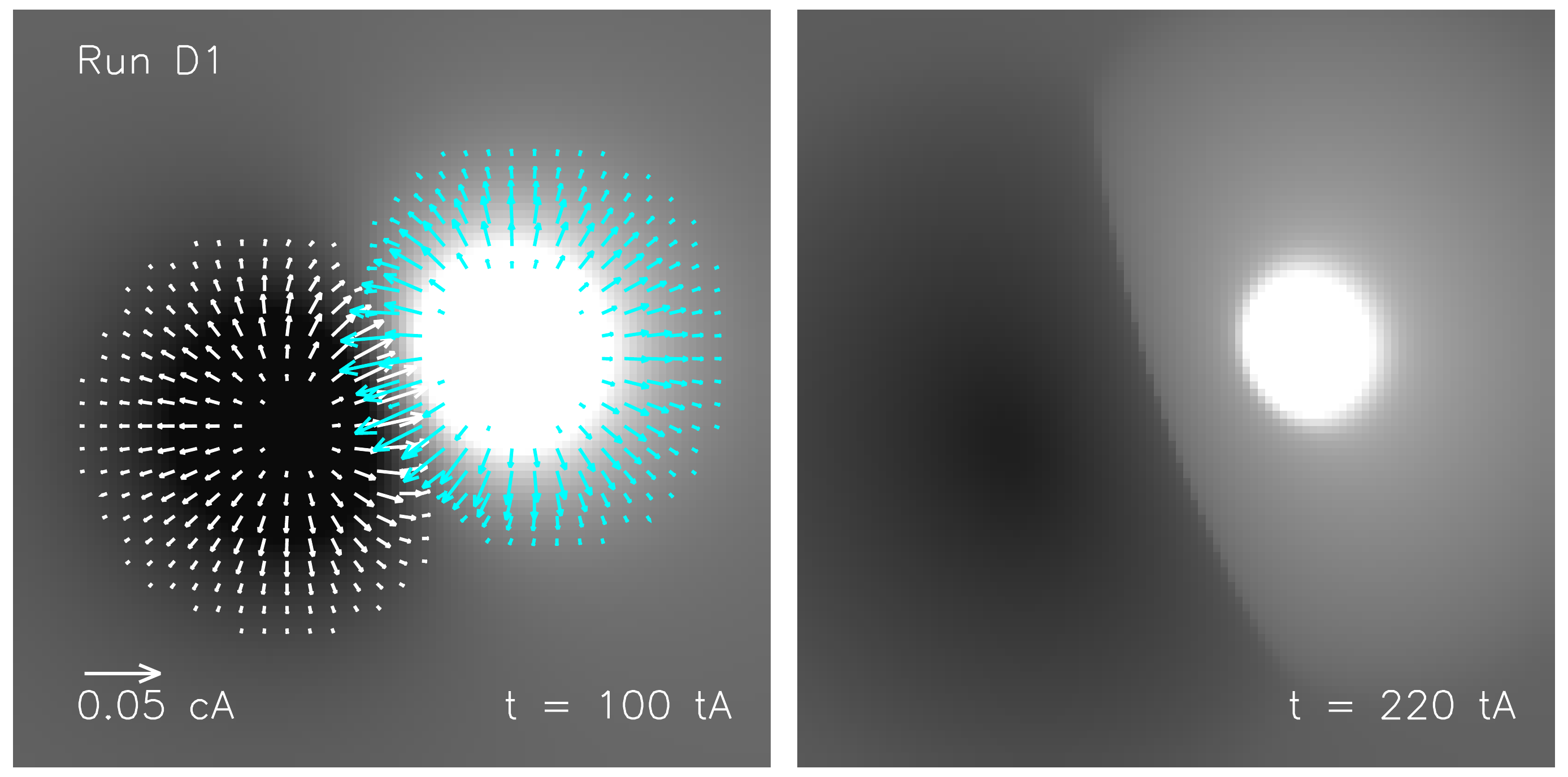}}
\subfigure{\label{RunD2-vfield}
\includegraphics[width=.47\textwidth,viewport=6 4 793 392,clip]{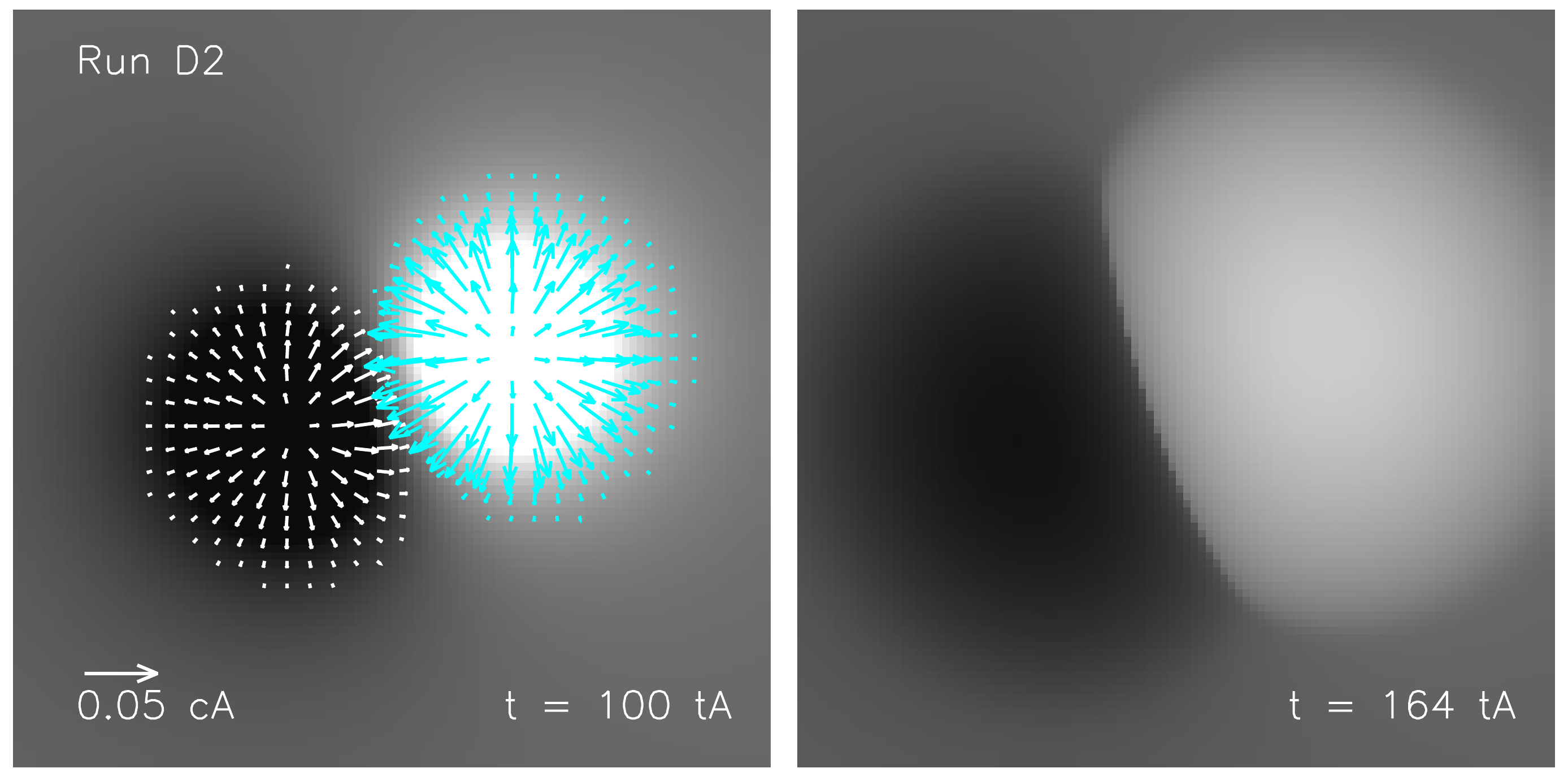}}
\caption{Maps of $ B_{z}(z=0)$ at the beginning/end (left/right columns) of the convergence phase for `Run~C', `Run~S', `Run D1' and `Run D2' (first, second, third and fourth row, respectively). White/black color indicate positive/negative magnetic field. Cyan/white arrows outline the initial velocity profile. The field of view is $ x,y \in [-3.3,3.3]$.}
\label{v-field}
\end{center}
\end{figure}

The different flow profiles, $\mathbf{u}_0(t)$, at the beginning of the convergence phase are shown in Figure~\ref{v-field} (left-column). The effect of these motions on the normal component of the magnetic field at $z=0$ and at the end of the driving phase is shown in  Figure~\ref{v-field} (right-column). 

A comparison between the last two rows of Figure~\ref{v-field} illustrates the effect of the parameter $\psi_1$ in the function $\psi(t)$: it controls the size of the region (centered around the maximum of the polarity) that is unaffected by the flows. Finally, all the applied flows have a component that induces flux convergence toward the PIL. This is a common features of several CME's initiation scenarios and it is very often observed to preceded the onset of eruptive flares.  

\begin{figure*}
\begin{center}
\includegraphics[width=.95\textwidth]{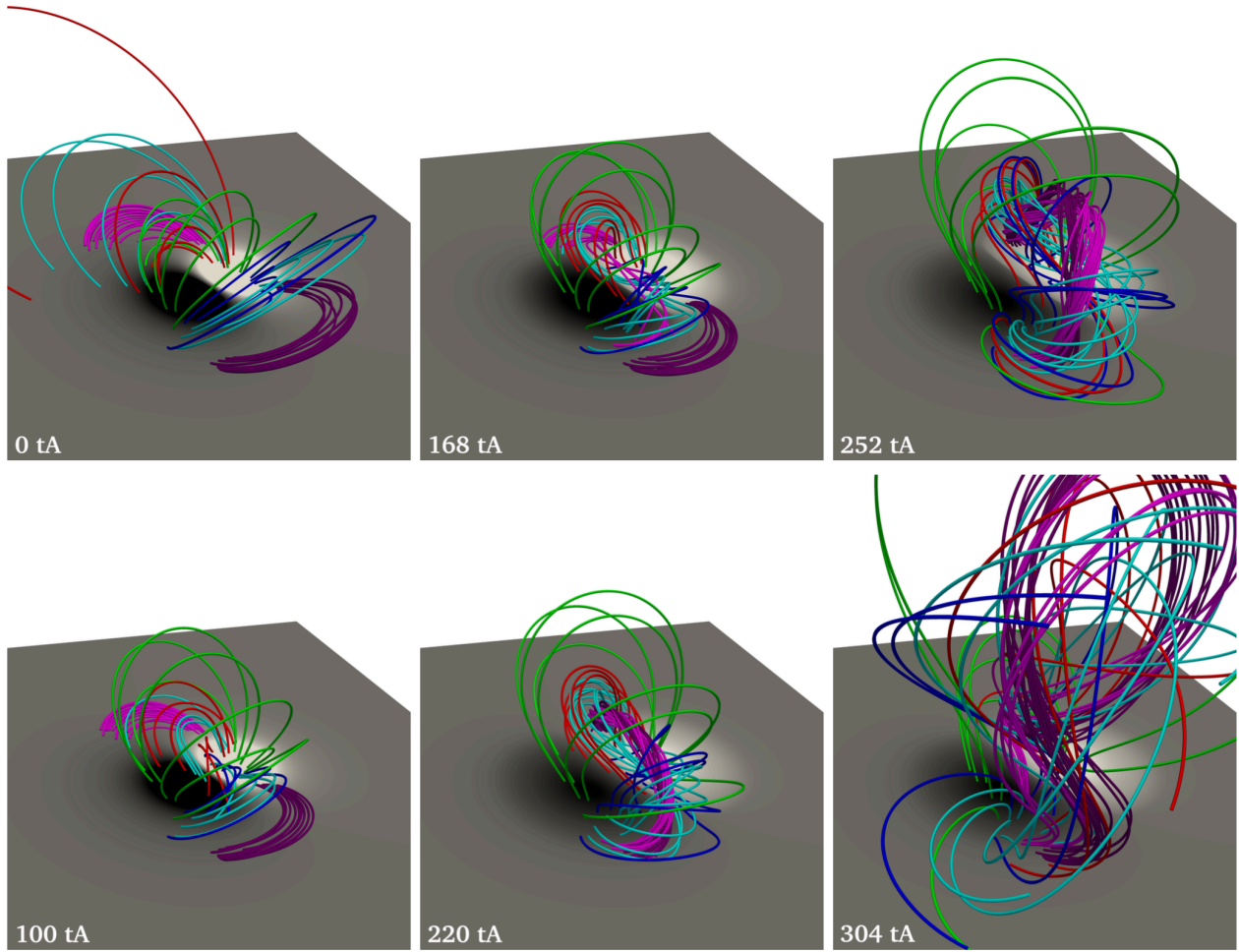}
\caption{Evolution of the magnetic field for `Run D1' during the quasi-static shearing/convergence phase (left/middle panels) and the eruptive phase (right panels). White/black color indicate positive/negative $B_{z}(z=0)$  } 
\label{3Devol}
\end{center}
\end{figure*}

\subsection{Diffusive coefficients} \label{diff_coeff}

The MHD equations solved by the OHM-MPI code include an artificial density diffusion coefficient $\xi$, a pseudo-viscosity $\nu'$ and a coronal resistivity $\eta$. These coefficients are required to ensure the numerical stability of the code.  In addition, at the lower boundary a photospheric diffusion term $\eta_{p}$ is also present.   

Each of the four simulation runs described in this paper can be divided into three different phases: (1) the twisting phase driven by the boundary motions defined by Equation~(\ref{twist}) and applied from $t=0$ to $t=100t_A$ (the same for all the simulations), (2) the convergence phase driven by the motions defined in Equations~(\ref{mC})-(\ref{mD2}) from $t=105t_A$ to $t=t_1$, i.e., the time when the eruption becomes unavoidable and (3) the eruption phase, where we impose $u_x(z=0,t)=u_y(z=0,t)=0$, that begins from $t=t_1$ till the end of the simulations. 

During the twisting and the convergence phases the coronal diffusivity is fixed to  $\eta_{\text{cor}} = 4.8 \times 10^{-4}$,  the diffusion coefficient  is $\xi = 1.5 \eta$ and the pseudo-viscosity is $\nu' = 25$.  At the smallest grid size these coefficients result in the following diffusive speeds $u_{\eta} =0.08$, $u_{\xi}=1.5u_{\eta}$ and $u_{\nu'}=0.15$. During the eruption phase, fast flows and sharp currents develop. Therefore,  we fix $\eta_{\text{cor}_1} = 2.1\times 10^{-3}$, $\nu'_1 = 41.7$ and $\xi_1 =  \eta_{\text{cor}_1}$. This is required to  ensure the numerical stability of the code.  

At the line-tied boundary, the photospheric resistivity  $\eta_{p}$ is set to zero during the twisting phase. This allows the build up of magnetic shear and current in the system without modifying the photospheric distribution of $B_z$ and without forming  bald-patches like in \cite{Aul2010}. During the convergence phase, magnetic flux is advected towards the PIL with a rate that varies with time in a nonlinear way, but that is always smaller then $2u_0^{\text{max}}$. Therefore, in order to avoid flux pileup at the PIL during the phase  $t \in [100t_A,t_1]$ we impose a photospheric resistivity $\eta_{p} =\eta_{\text{cor}}$. We choose this value because it seems to be the best compromise between magnetic field diffusion and flux advection for a significant portion of the PIL.   Finally, during the eruption phase ($t \geq t_1$) the photospheric diffusion is again set  to zero. This eventually results in the onset of a numerical instability close to the boundary at $z=0$ that originates from narrow photospheric  current layers. This problem is resolved by artificially smoothing the Lorentz force within the first twenty-two grid points, i.e., for $z \in [0,0.165]$. This is achieved by multiplying the Lorentz force in the momentum equation by a factor 

\begin{equation}
 \sigma_f= \frac{1+\tanh \left[33(z-0.08)\right]}{2},  ~~~~~ \forall z \in [0,0.165].
\end{equation}

We choose this solution instead of an increased photospheric diffusion, because this latter would induce reconnection at the line-tied boundary and, more importantly, will essentially modify the reference potential field energy during the study of the instability.

\section{Dynamics and Energetics} \label{Evolution}

In this section the evolution of the system from the initial current-free configuration to the formation of the flux rope and its eruption is discussed. Although the timing and the onset of the eruption differs between the simulations, the  overall topological evolution of the system is the same for the different runs. Therefore, we limit ourselves to the description of only one of the four simulations, namely `Run D1', that is, a partial dispersal of the active region field. 

\subsection{Dynamical Evolution} \label{Evolution1}

Figure~\ref{3Devol} (top-left) shows selected magnetic field lines highlighting the initial potential magnetic field configuration of the system. In order to build up currents in the system at the line-tied boundary we apply the velocity field defined in Equation~(\ref{twist}).  As already mentioned, the applied motions are never larger then few percents of the Alfv\'{e}n speed and, as a consequence, the coronal field evolves quasi-statically in response to the photospheric driver.  

The configuration of the system after $100t_A$, i.e., at the end of the twisting phase, is shown in Figure~\ref{3Devol} (bottom-left). As evident the field lines close to the PIL (blue/red/cyan field lines) are the most sheared ones, while the overlying field (green field lines) only experiences a minor twist. During this phase no magnetic flux rope is observed. Nevertheless, due to the injected shear, the magnetic pressure starts to increase, especially in the proximity of the PIL, and the system slightly bulges up. In the simulation of \cite{Aul2010} this initial bulging up of the field lines close to the PIL combined with the presence of a finite photopsheric diffusion resulted in a topological change of the system from sheared-arcade to a three-dimensional bald-patches configuration containing a bald-patch separatrix. Because we explicitly impose $\eta_p=0$, no flux cancellation occurs close to the PIL and no change in the topology is observed during the twisting phase in our simulations, hence no magnetic flux rope is formed yet.

The aim of this first driving phase is to obtain a quasi-NLFF field with a sheared-arcade topology. This constitutes the initial condition for our study of different classes of photospheric motions. These latter induce a deformation and evolution of the active region similar to what is often observed on the Sun. 

Starting from this sheared-arcade configuration we apply the four different boundary motions described in Equations~(\ref{mC})-(\ref{mD2}). Figure~\ref{3Devol} (middle panels) shows the response of the system to these drivers (for `Run D1'). 

As a consequence of the applied convergence motions photospheric flux is advected toward the PIL and, since during this phase we impose a finite photopsheric diffusion, the three components of the field at the PIL are canceled and, similarly to \cite{Aul2010}, we observe the transition from sheared-arcade to a bald-patch topology. Magnetic reconnection  occurs at the bald-patch and a mildly twisted flux rope is formed. 

Figure~\ref{3Devol} (middle-top) shows the configuration of the system during the build up phase of the flux rope. The forming magnetic flux rope is highlighted by selected pink/purple/cyan field lines. Pink (purple) field lines are traced starting from the top (bottom) part of the positive (negative) polarity. Cyan field lines are traced starting from locations that are in between the pink/purple and the red/blue field lines. The red/blue field lines are traced starting at either side of the bald-patch actually highlighting the bald-patch separatrix below the magnetic flux rope. The figure also shows that the flux rope field lines are about twice as long as the field lines that extend at either sides of the bald-patch separatrix. 

While the convergence motions continue, more and more flux is advected towards and canceled at the PIL, and the bald-patch separatrix evolves. In particular, the photospheric footprints of the bald-patch separatrix continue to expand. Figure~\ref{3Devol} (middle-bottom) shows a snapshot of the system towards the end of this process and just before the onset of the instability (see Section~\ref{TI}). The purple and pink field lines show the main body of the 
flux rope. The cyan field lines show the outer layer of the flux rope, that is, the set of field lines belonging to the flux rope that come closest to the bald-patch. Due to the curvature of the flux-rope axis, only those field lines in the outer layer contain magnetic dips, while those close to the axis do not. This may be contrasted with an ideal cylindrical flux rope, where dips are present at all distances from the axis. The figure also shows how the flux rope is embedded between two set of sheared field lines (red/blue) and much less sheared (i.e., quasi-potential) overlying magnetic field (green field lines). 

The formation phase of the magnetic flux rope is morphologically similar --- but topologically different --- to what was observed in the simulation of \cite{Aul2010}. In the latter, the bald-patch separatrix disappeared quite soon during the formation phase of the magnetic flux rope and was replaced by a quasi-separatrix layer. The reconnection that transferred flux from the overlying field to the magnetic flux rope occurred at the hyperbolic flux tube. However, in our simulations, the bald-patch separatrix is always present and the magnetic reconnection occurs at the bald-patch only. 

Figure~\ref{3Devol} (right-top) shows the early stages of the flux rope eruption. During this phase photospheric diffusion is reset to zero and no boundary flows are applied. Therefore, the system only evolves in response to the imbalance between the magnetic pressure associate with the current carrying magnetic flux rope and the magnetic tension of the overlying field. Furthermore, during the early stages of the eruption, the flux rope expands outwards without any significant kink. 
When the flux rope enters the dynamical regime a current sheet is formed in the corona initially above the bald-patch. Below the current sheet sheared post-flares loops similarly to \cite{Aul2012} are formed (not displayed in the figure). During this phase bald-patches are still present and bald-patches field lines are now low-lying field lines with the second foot point anchored close to the central part of the PIL. 

The continuous presence of bald-patches in our simulations \citep[in contrast with][]{Aul2010} is probably due to the convergence motions. As a consequence of these motions, magnetic field is continuously advected towards the PIL at rate that is comparable to the diffusion rate, and this maintains the bald-patch topology.
   
The final snapshot of the simulation is shown in Figure~\ref{3Devol} (bottom-right), just before the fast flows generated during the eruptive phase induces a numerical instability and halts the simulation.  The flux rope undergoes a full eruption and, during the propagation, is deflected towards the bottom-left part of the domain. This deflection is probably a consequence of the asymmetry of the system. The positive polarity is more intense than the negative one, which results in a magnetic pressure gradient that is directed towards the bottom-left part of the domain. This kind of flux rope deflection has also been found in a simulation of a more complex, asymmetric active region \citep{Zuc2012c}.

\subsection{Energy evolution} \label{energy}

\begin{figure}
\begin{center}
\includegraphics[width=.5\textwidth,viewport= 10 0 550 390]{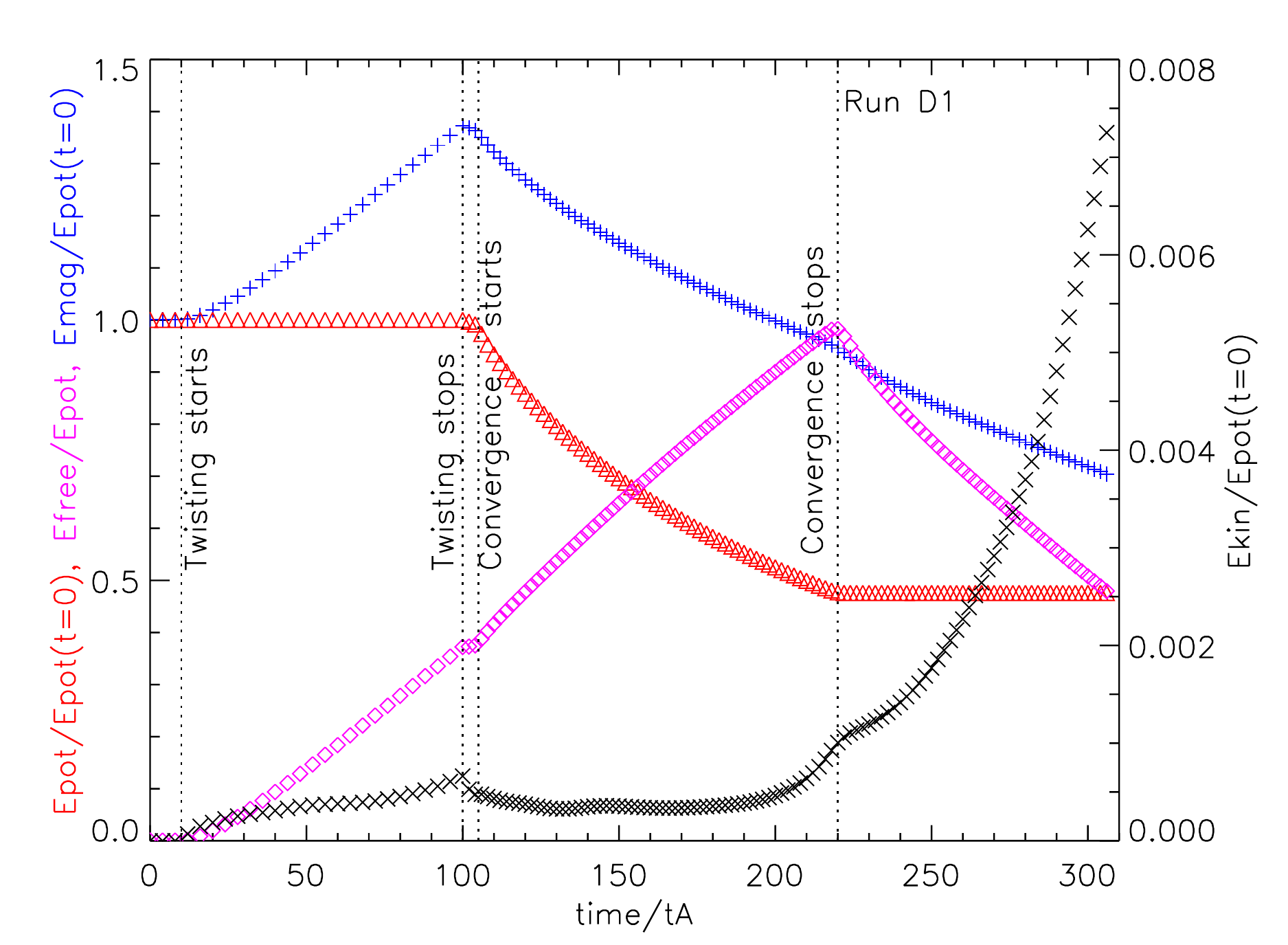}
\caption{Temporal evolution of the free magnetic energy in the system (purple `$\diamond$') as well as of the kinetic (black `$\times$'), potential (red `$\bigtriangleup$') and total (blue `$+$') magnetic energy of the system normalized to the energy of the initial (potential) magnetic field for  `Run D1'. The magnetic energy of the initial potential field is $E_{\text{pot}}(t=0)=174.4$. }
\label{RunD1-energy}
\end{center}
\end{figure}

\begin{figure*}
\begin{center}
\subfigure{\label{RunC-currents}
\includegraphics[width=.45\textwidth,viewport= -30 -10 420 400]{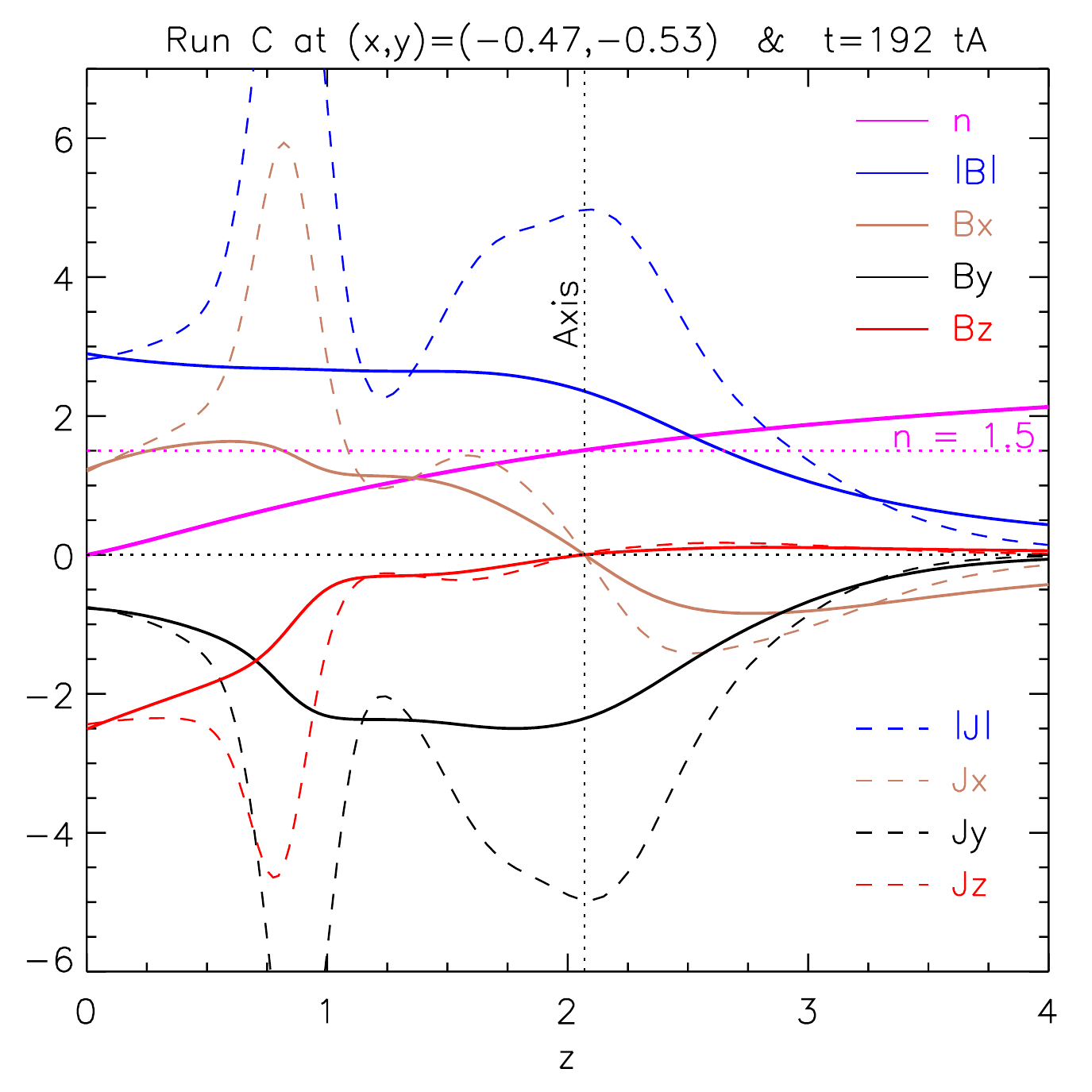}}
\subfigure{\label{RunS-currents}
\includegraphics[width=.45\textwidth,viewport= -30 -10 420 400]{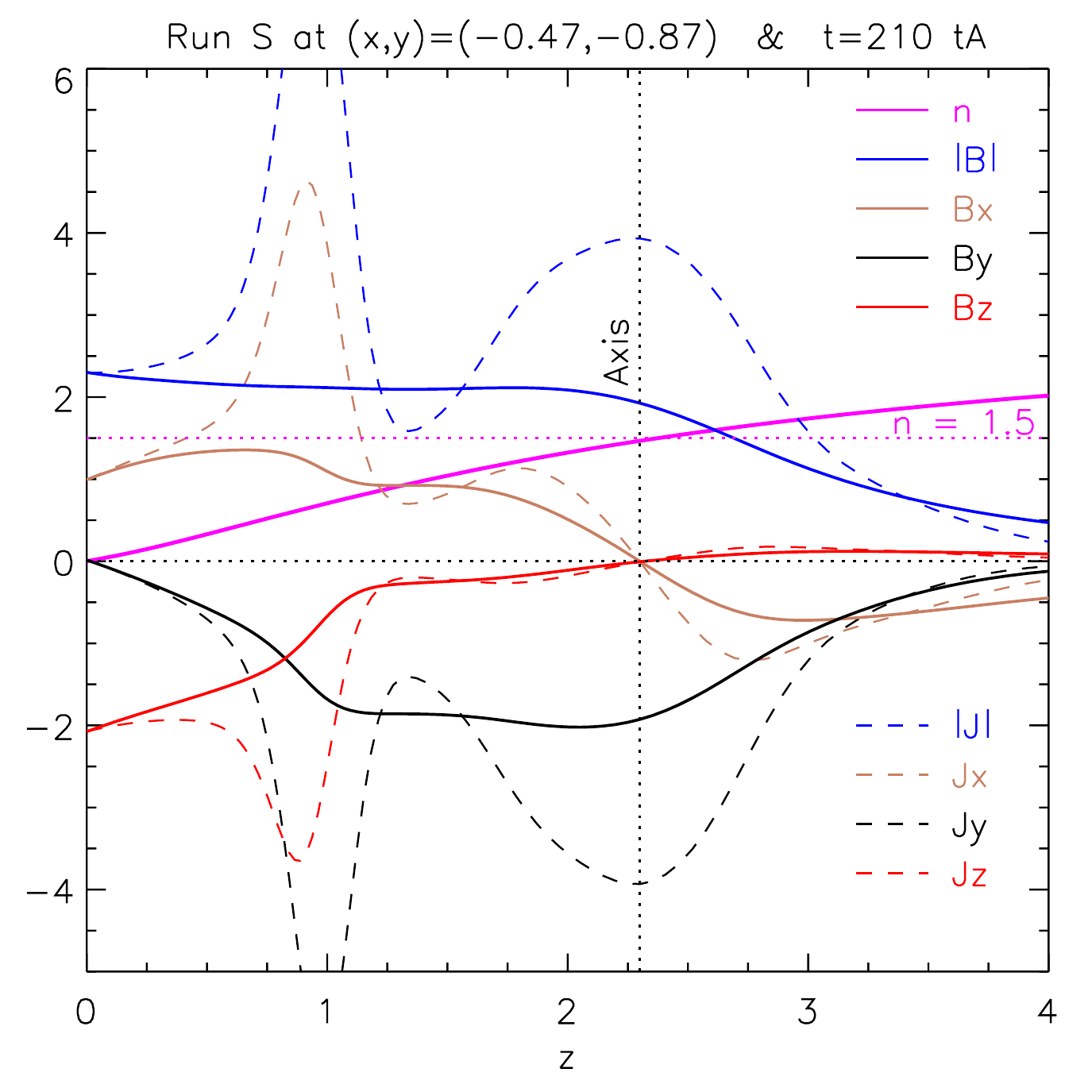}}
\subfigure{\label{RunD1-currents}
\includegraphics[width=.45\textwidth,viewport= -30 -10 420 400]{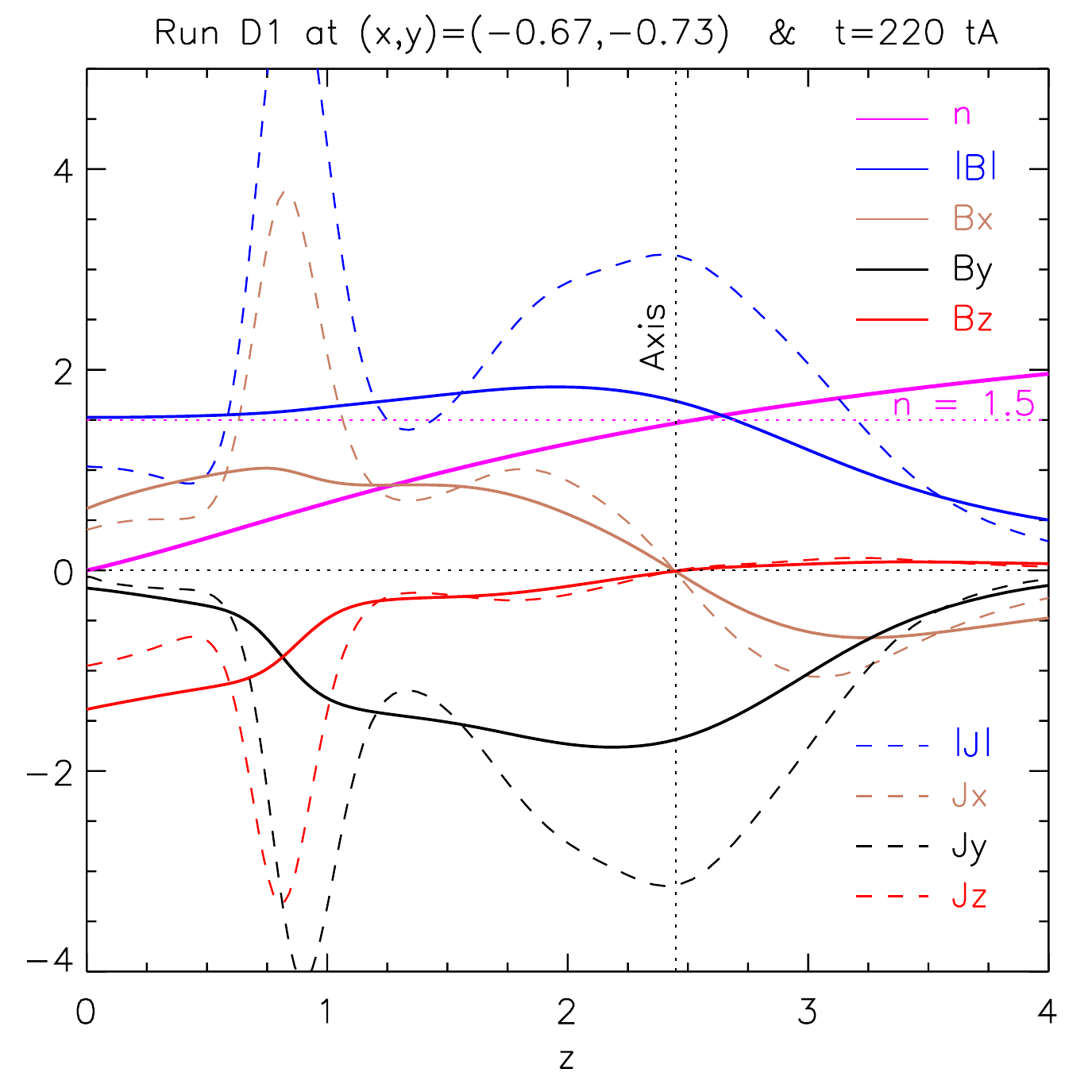}}
\subfigure{\label{RunD2-currents}
\includegraphics[width=.45\textwidth,viewport= -30 -10 420 400]{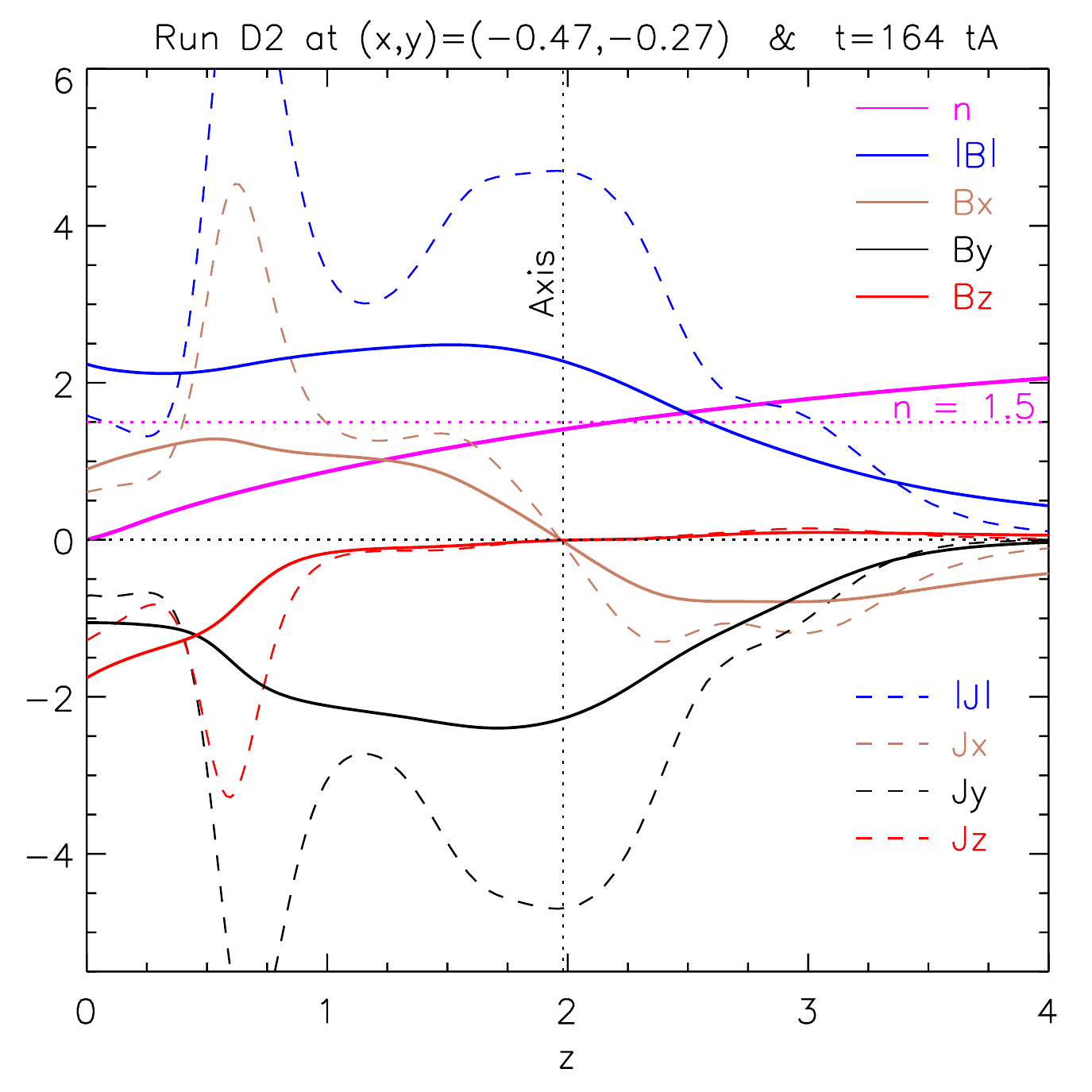}}
\caption{Vertical dependence of the norm, $x$, $y$ and $z$ components (blue, brown, black and red lines) of the magnetic field/current density (solid/dashed lines) as well as of the decay index (solid purple line) at a given (x,y) position and time $t$ (indicated in each panel) for `Run~C', `Run~S', `Run D1' and `Run D2' (from top-left to bottom-right, respectively). The vertical dotted line indicates the height where both B$_{x}$ and B$_{z}$ change sign.}
\label{Currents}
\end{center}
\end{figure*}

\begin{figure*}
\begin{center}
\includegraphics[width=.95\textwidth,viewport= 0 100 600 720,clip]{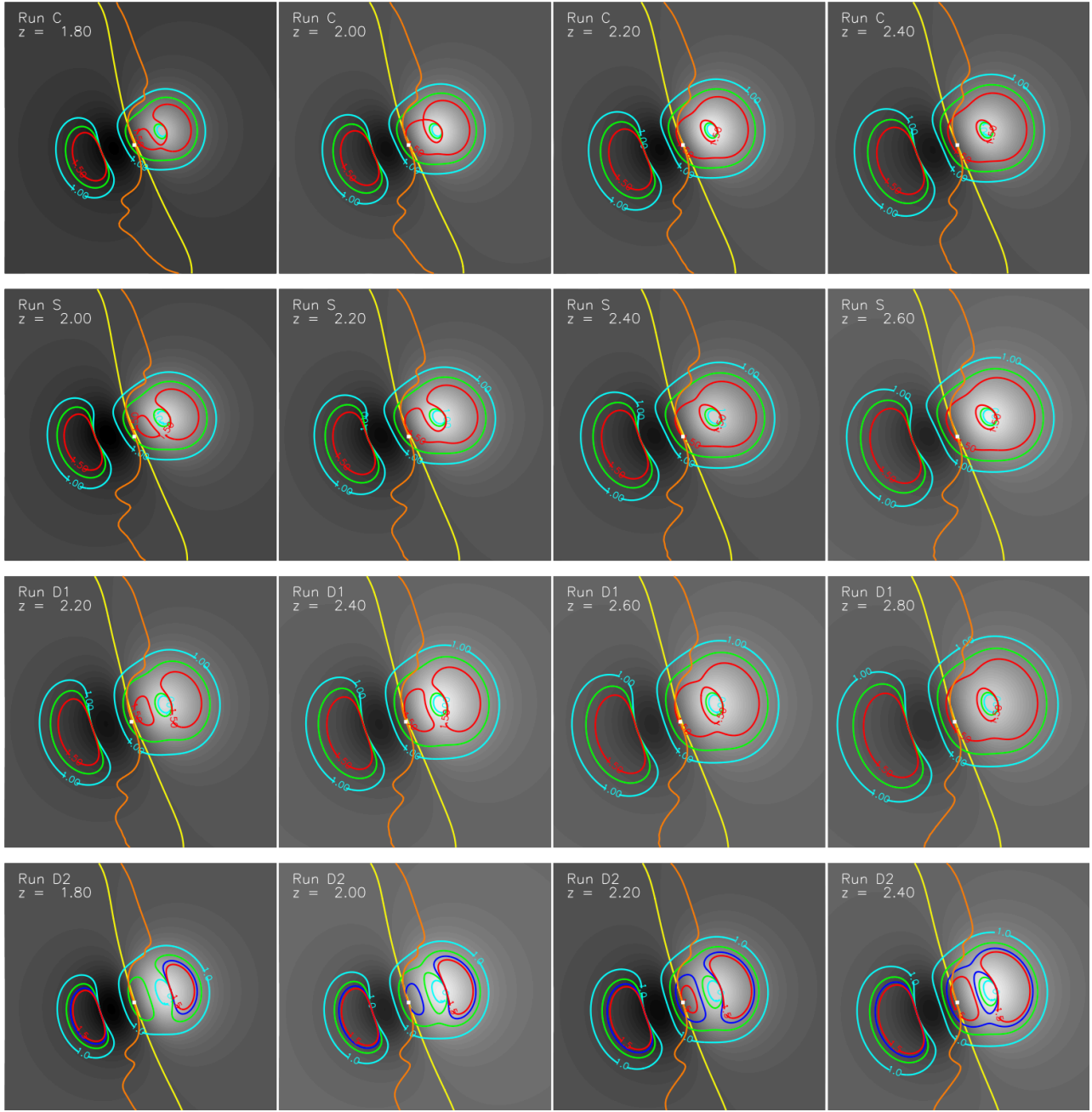}
\caption{Two dimensional maps of contours of the decay index --- $n=1$ (cyan), $n=1.25$ (green), $n=1.5$ (red) --- over plotted to the normal component of the magnetic field at different heights for `Run~C', `Run~S', `Run D1' and `Run D2' (first, second, third and fourth row, respectively). For 'Run D2' the contour of the decay index at $n=1.4$ (dark blue) is also shown. The orange (yellow) line indicate the PIL of the simulated (extrapolated) magnetic field. The white square indicates the (x;y) position of the apex of the flux rope's axes. The extrapolations have been performed at times $194t_A,~214t_A,~220t_A$ and $164t_A$ for `Run C, S, D1' and 'Run D2', respectively.}
\label{Decay}
\end{center}
\end{figure*}

\begin{figure*}
\begin{center}
\includegraphics[width=.95\textwidth,viewport= 90 0 700 630,clip]{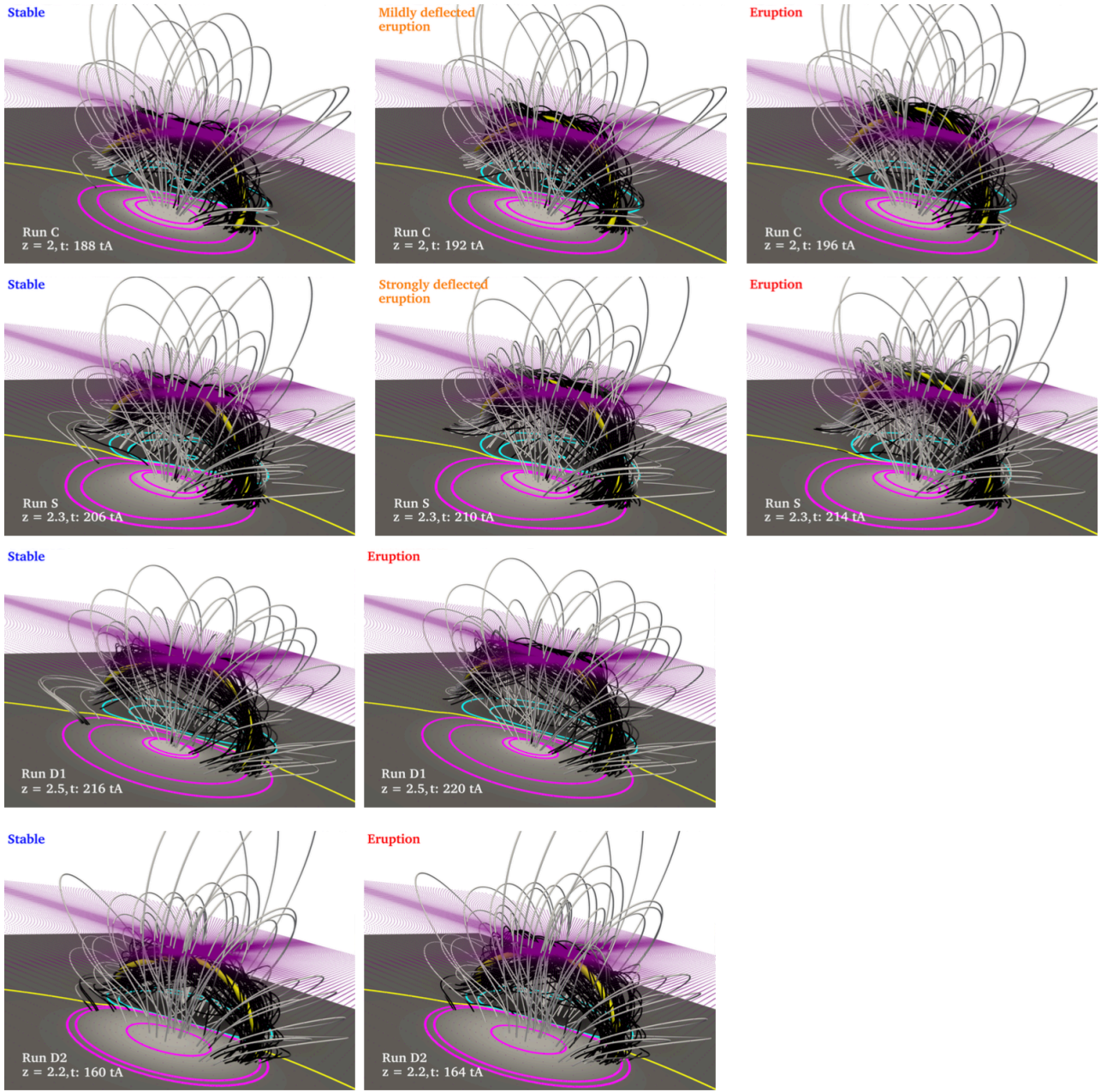}
\caption{Snapshots around the time of the instability onset for `Run~C', `Run~S', `Run D1' and `Run D2' (first, second, third and fourth row, respectively). The purple semitransparent plane indicate the height ($z$) --- as deduced from Fig.~\ref{Decay}--- where the contour of the decay index, $n=1.5$, touches the part of the PIL occupied by the flux rope.  The field lines are color coded with the magnitude of the current density using the same difference between the low and upper limits of the color scale.   } 
\label{3DView}
\end{center}
\end{figure*}

\begin{figure*}
\begin{center}
\includegraphics[width=.95\textwidth,viewport= 10 200 750 420,clip]{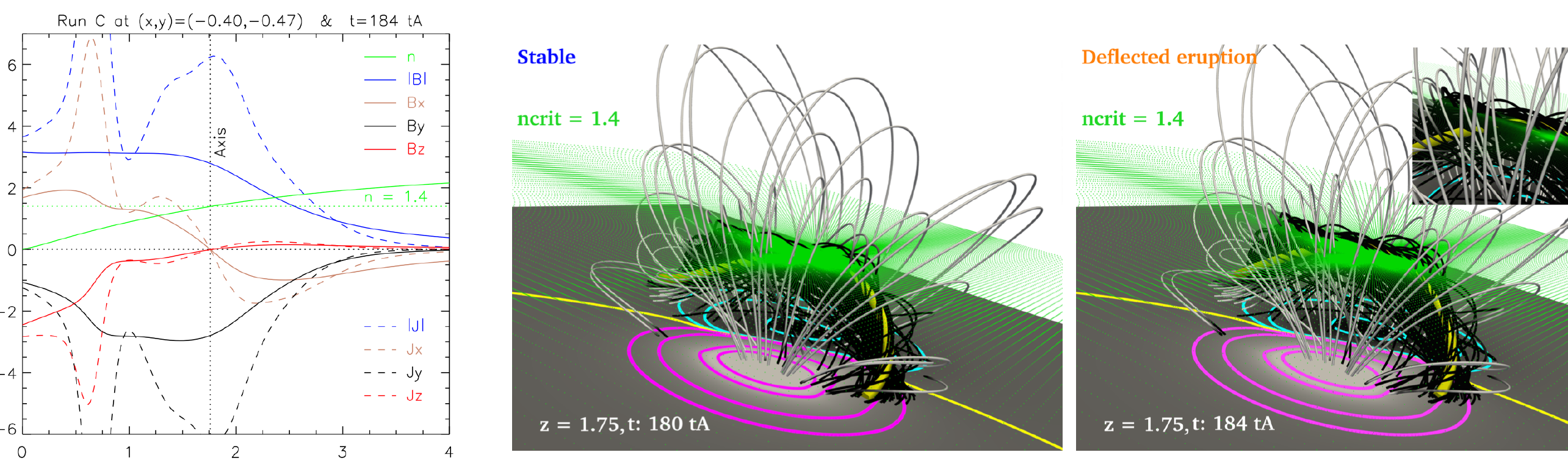}
\caption{ \textit{Left panel:} Same as Figure~\ref{Currents} for `Run~C' under `low' coronal diffusion conditions. \textit{Right Panels:}  Two snapshots around the time of the instability onset for `Run~C' under `low' coronal diffusion conditions. The green semitransparent plane indicate the height ($z$) where the critical decay index at the apex of the magnetic flux rope is $n_{crit}=1.4$.  The field lines are color coded with the magnitude of the current density and the yellow thick tube indicates the magnetic flux rope's axis.}
\label{Current-test}
\end{center}
\end{figure*}

The evolution of the potential, free and total magnetic energies as well as of the kinetic energy for `Run D1' is shown in Figure~\ref{RunD1-energy}. The same plots for the other simulation runs are presented in Appendix~A. 

To compute the potential energy of the system for each snapshot we perform a potential-field extrapolation of the photospheric magnetic field $B_z(z=0)$ using the method of \cite{Ali1981}. First, we extract the bottom boundary of the simulation and remap the non-uniform grid of OHM-MPI onto a uniform grid of $[301 \times 301]$ points resulting in a uniform grid spacing of about ten times the smallest OHM-MPI grid cell. Second, in order to minimize the aliasing effects due to the intrinsically assumed periodic boundary of the FFT method, the remapped OHM-MPI boundary is inserted at the center of a eight times larger grid that is padded with zeros. With this method we achieve an accuracy in the magnetic energy computation, i.e., the difference between the energy of the reconstructed field and the energy of the field generated by the analytic charges of about  0.1\%.  

Due to the nature of the twisting flows ---which do not modify $B_z$ at the boundary from $t=10t_A$ to $t=100t_A$--- the potential energy of the system (`$\bigtriangleup$'~signs) remains unchanged and the increase in the total magnetic energy (`$+$'~signs) is directly related to the increase of the free magnetic energy (`$\diamond$'~signs). However, during the convergence phase, i.e., from $t=105t_A$ to $t=220t_A$, the distribution of $B_z$ at the boundary is modified by the flows and the potential energy of the system changes, eventually reaching about half of its initial value. During this phase, the total magnetic energy also decreases, however, the free magnetic energy continues to increase. At the end of the convergence phase, the system has a free magnetic energy that is of the same order of the potential energy, therefore, about half of the total magnetic energy stored into the system is available for the eruption. 

During the twisting and convergence phases, the kinetic energy always remains very low, eventually confirming that the system evolves quasi-statically. At about $t=210t_A$ the kinetic energy starts to rise exponentially, but at $t=220t_A$ the pseudo-viscosity is increased (Section~\ref{diff_coeff}) and the initial rise of the kinetic energy is smoothed out for about $5-10t_A$. After that, the exponential increase of the kinetic energy continues until the end of the simulation. As a final remark we  note that the kinetic energy constitutes only a very moderate fraction ($\sim 4 \%$) of the free magnetic energy released during the eruption.

\section{ Eruption onset and trigger}\label{TI}

The dynamical evolution of the system and the previous results of \cite{Aul2010} suggest that the torus instability is the trigger of the flux rope eruptions presented in this paper. Moreover, as discussed by \cite{Dem2010}, the exact value of the critical decay index at the onset of the eruption may be different depending on the exact morphology of the flux rope. 

To clarify these points, in this section we present the analysis that we performed in order to determine: (1) the value of the critical decay index at the onset of the eruption and (2) if the torus instability is the  trigger of our eruptions. In particular, we first determine the moment of the eruption (i.e., $t_1$) by performing relaxation runs, then, around the moment of the eruption, we identify the axis of the flux rope and compute the decay index at different heights. Finally, we determine the value of the decay index at the height of the flux rope axis for both eruptive and non-eruptive runs.

In the following subsections the different stages of our analysis are described in details. 

\subsection{Relaxation Runs} \label{relax}

In order to determine the time of the onset of the eruptions, at different times $t^{\ast}$  during the convergence phase we impose $\mathbf{u}_0(t \geq t^{\ast}) =0$, $\eta_p =0 $ and let the system evolve. The photsopheric flows are always slowed down by using the function $\gamma(t)$ defined in Equation~(\ref{gamma}). The first time $t^{\ast}$ for which the eruption becomes unavoidable defines the time $t_1$ discussed in Section \ref{boundary} and Equation~(\ref{gamma}).

For `Run C' we find that for $t^{\ast}=188t_A$ no eruption occurs, while for $t^{\ast}=192t_A$ the flux rope erupts, but it gets deflected. In fact, at the initial phase of the instability onset the magnetic pressure of the magnetic flux rope is just enough to overtake the magnetic tension of the overlying field and the flux rope starts to ascend. Due to the asymmetry of the configuration, the magnetic field of the positive polarity is larger than the one of the negative polarity and a magnetic pressure gradient exists. During the early stages of the eruption, this pressure gradient influences the dynamic of the flux rope, eventually deflecting it towards the bottom-left boundary. If the photospheric motions for `Run C' are stopped at $t^{\ast}=196t_A$ the flux rope experiences a `full eruption', like the one discussed in Section~\ref{Evolution}. During this extra time, extra current-carrying magnetic flux (and hence magnetic pressure) is injected into the flux rope, eventually mitigating the effect of the asymmetric pressure gradient due to the external field. 

It should be noted that the deflection of the magnetic flux rope is observed also in the `fully eruptive' runs (Section~\ref{Evolution}), but develops at a later stage. For the fully eruptive runs the pressure gradient associated with the flux imbalance influences only the propagation of the magnetic flux rope rather than its initial dynamics. 

For `Run S' we observe a  similar behavior. If we stop the phostospheric driver at $t^{\ast}=206t_A$, the system finds a new equilibrium, while if the photospheric driver is switched off at $t^{\ast}=210t_A$, an eruption occurs, but undergoes a very strong deflection towards the boundary. If $t^{\ast}=214t_A$, a full eruption occurs. 

For `Run D1' (`Run D2', respectively) the evolution is slightly different. If the photospheric driver is interrupted at $t^{\ast}=216t_A$ ($t^{\ast}=160t_A$, respectively) no eruption occurs, but if the driving is stopped four Alfv\'{e}n times later, i.e., at $t^{\ast}=220t_A$ ($t^{\ast}=164t_A$, respectively), the system undergoes a full eruption without displaying the deflection that is observed for `Run C' and `Run S'. One of the differences between `Run C, S' and `Run D1, D2' is that in the latter the prescribed motions are perpendicular to the PIL, while the former ones are parallel to the $x$-axis inducing a photospheric flux distribution that eventually increases the original asymmetry (and magnetic pressure imbalance) of the overlying magnetic field.      

\subsection{Flux rope axis} \label{axis}

Magnetic flux ropes are generally defined as an ensemble of twisted magnetic field lines that wrap around a common central axis. However, when the twist is  $\lesssim 2\pi$ the wrapping is not full and the identification of the axis of the magnetic flux rope is difficult. As a consequence, a unique criterion for the identification of the axis of a magnetic flux rope does not exist and different cases should be analyzed individually. 

For curved flux ropes that are in equilibrium with an external magnetic field, the apex of the flux rope's axis must lie along the local PIL. Therefore, at the location of the apex, $B_z$ must change sign. Moreover, in our simulations the flux rope (at least the highest part of it) is almost parallel to the $y$-axis, therefore, at the height of the axis, $B_x$ must changes sign with the height $z$. 

To determine the axis of the magnetic flux rope we then proceed as follow. For the simulation output at time $t=t_1$ ($192t_A,210t_A,220t_A,164t_A$ for  `Run C, S, D1, D2', respectively), we look at the three components of the magnetic field along vertical lines passing through different $[x;y]$ positions along the PIL. Among the different positions, we look for the one $[\bar{x};\bar{y}]$  where the change in sign of $B_x$ and $B_z$ occurs at the same height $z=\bar{z}$. The axis of the magnetic flux rope (see yellow tube in Figure~\ref{3DView}) is then assumed to be the  magnetic field line that passes through the position $[\bar{x};\bar{y};\bar{z}]$ at time $t=t_1$. 

Figure~\ref{Currents} shows the norm and the three components of the magnetic field (solid lines) and of the current density (dashed lines) along the vertical $z$-axis that passes through $[\bar{x};\bar{y}]$. 

The vertical dotted line indicates the height $\bar{z}$ where both $B_x$ and $B_z$ change sign, that is, the height of the apex of the magnetic flux rope's axis. As evident the apex of the magnetic flux rope does not have the same position in all the simulations. In particular, for `Run C' and `Run D2' the eruption begins at a lower heights than in  `Run S' and `Run D1'. 

Interestingly, we find a posteriori in Figure~\ref{Currents} that the current density has a local maximum along the axis of the magnetic flux rope  and is almost aligned with it ($J \simeq |J_y|$). This behavior is quite similar to what is expected for a flux rope generated by a relatively thick current wire or a NLFF cylindrical constant-twist (Gold-Hoyle) flux rope. 
Closer to the photosphere the maximum of the norm of the current density  highlights the narrow current layer associated with the bald-patch separatrix. This is the region where magnetic reconnection occurs eventually transferring flux from the overlying filed into flux rope field.

\subsection{Decay index}\label{decay-index}

The magnetic field that is relevant to compute the decay index is the so called `external field' that is not associated with the coronal currents. In the symmetric configurations where the torus instability has been originally formulated the `external field' is the one generated by the sub-photospheric charges. At the axis of the flux rope, it has only a toroidal component.   

We use the potential field associated with the photospheric distribution of  $B_z(x,y,z=0)$ at time $t=t_1$ (see Section~\ref{energy}) as the external field to be used for the computation of the decay index. Furthermore, to account for the intrinsic asymmetry of our model only the horizontal component of the reconstructed potential field is used.

Two dimensional maps of the decay index for different heights and for the four different simulations are shown in Figure~\ref{Decay}. Only the contours where $n=1,1.25$ and $1.5$ are shown (for `Run D2' an extra contour at $n=1.4$ is also included). The yellow (respectively orange) line indicates the PIL of the potential field (respectively in the simulation) at any given height. Finally, the white square indicate the $[\bar{x}, \bar{y}]$ position of the apex of the magnetic flux rope's as determined in Section~\ref{axis}. 

Figure~\ref{Decay} (first row) shows maps of decay index for `Run C'. The contour $n=1.5$ touches the local PIL of the simulation (i.e. the PIL of the flux rope) at a height $z\simeq 2$. This is remarkably close to the height of the axis of the magnetic flux rope at $t=192t_A$ (Figure~\ref{Currents}, top left), when the deflected eruption began.   

For `Run~S'  at the onset of the eruption, the axis of the magnetic flux rope has a height $z\simeq 2.3$. Figure~\ref{Decay} (second row) shows that the critical value $n=1.5$ of the decay index at the location of the apex of the magnetic flux ropes's axis is reached between $z=2.2$ and $z=2.4$. This is in very good agreement with the original $n_{crit} =1.5$ of the torus instability scenario. A  similar conclusion can be drawn also for `Run D1', where the actual value of the decay index at the position of the apex is $n_{crit} \simeq 1.45$ (Figure~\ref{Currents}, bottom-left and Figure~\ref{Decay}, third row). 


From Figure~\ref{Currents} (bottom right) we can deduce that for `Run~D2' the eruption begins when the flux rope axis has a height $z\simeq 1.95$. As evident the decay index at the apex of the magnetic flux rope's axis (purple curve) has not yet reached the value $n=1.5$. This is more evident in Figure~\ref{Decay} (fourth row): at $z=2$ the contour $n=1.4$ has already touched the PIL, while the  contour $n=1.5$ touches the PIL only at $z=2.2$. This seems to suggest that for `Run~D2' the critical decay index for the onset of the eruption is about 7\%  smaller than for the other three cases. 

\subsection{Unique critical decay index ?}

The results of the performed analysis are synthesized in Figure~\ref{3DView}.  In this figure for the simulation runs that exhibited an initial deflection, i.e., `Run C' and `Run S', three different snapshots around the moment of the onset of the eruption are presented. For `Run~D1' and `Run~D2' no deflected eruption is observed and therefore only two snapshots around the onset time are shown.  

For each snapshot of the same row (i.e., same boundary flow profile), the magnetic field lines are traced starting from the same foot points and are color-coded with the current density. The thicker yellow field line represent the axis of the magnetic flux rope as identified in Section~\ref{axis}. The purple semitransparent plane represent the critical height (different for each simulation) where the contour $n=1.5$ of the decay index touches the PIL of the simulation. This height is $z=2,~2.3,~2.5$ and $z=2.2$ for Run C, S, D1 and D2, respectively (cfr. Section~\ref{decay-index} and Figure~\ref{Decay}). 

For each of the simulations the configuration displayed on the left-columns is stable, i.e., no eruption occurs if the photospheric driver is stopped (see Section~\ref{relax}). It is evident from Figure~\ref{3DView} (left-columns) that at this time the axis of the magnetic flux rope has not yet reached the height where $n=1.5$ and no eruption occurs. 

For the two deflected eruptions, i.e., `Run C' and `Run S', the middle-columns of Figure~\ref{3DView} (top two rows) report the configuration of the system when the deflected eruptions occur. As evident the axis of the flux rope has reached the height where $n=1.5$, and even if we stop the driver now the eruption  occurs anyway. The system has entered an unstable regime and evolves driven by the imbalance between the magnetic pressure of the flux rope and the magnetic tension of the overlying field. It is very interesting that for these two simulations the critical value for the onset of the torus instability is remarkably close to the theoretical value for a thin circular current ring \citep{Dem2010}. Finally, if we switch-off the flows four Alv\'{e}n times later (Figure~\ref{3DView} two top-right columns), the axis of the magnetic flux rope is well above the theoretical critical height and the eruption develops without any deflection (apart from the interaction with the boundary during the propagation). As already mentioned in Section~\ref{relax} this is probably due to the extra magnetic pressure built into the flux rope during the extra convergence time. 

Figure~\ref{3DView} (two bottom-right panels) show the eruptive configuration for `Run D1' and `Run D2'. As evident for both simulations the axis of the magnetic flux rope has not yet reached the height where $n=1.5$ (semitransparent purple plane), but the eruption occurs anyway. As can be deduced from the previous section, the value of the decay index is $n_{crit} \simeq 1.4-1.45$, that is, still very close to $n=1.5$. 

\subsection{The effect of the diffusion coefficients}\label{effect-diff}



During the eruption, strong flows originate at the current sheet that develops below the erupting flux rope, eventually resulting in a numerical instability and halting the simulation. As discussed in Section~\ref{diff_coeff} this can be resolved by increasing the diffusion coefficients. In order to minimize the variation between the different runs, the relaxation runs are performed with the same conditions as the eruptive ones. In other words, at time $t=t_{1}$ not only are the flows  switched off, but the coronal diffusive coefficients are also modified according to what is discussed in Section~\ref{diff_coeff}. This results in a overestimation of the critical value of the decay index. 

The system approaches the critical point for the onset of the instability in an environment that is characteristic by `low' diffusion coefficients. Under this condition, current is built up into the magnetic flux rope through magnetic reconnection at the bald-patch and the flux rope slowly rises. At a certain moment, it reaches the height where $n=n_{crit}$ and starts to accelerate under the effect of the torus instability. However, at the same time, we increase the coronal diffusivity by a factor of $4.37$, eventually suddenly dissipating part of the magnetic flux rope's current and bringing the flux rope back to the equilibrium curve. Therefore, in order for the eruption to occur, extra-current, behind the amount required under `low' diffusion conditions, must be injected into the flux rope to account for the increased diffusivity during the early stages of the eruption. During this extra time the flux rope quasi-statically raises a little further up, eventually leading to a slight overestimation of the critical value of the decay index.


For one selected run (`Run C') we performed some relaxation runs without increasing the coronal diffusive coefficients. Figure~\ref{Current-test} (left panel) shows the current distribution as well as the decay index along a vertical line passing through the $[\bar{x},\bar{y}]$ location of the apex.  In the `low' diffusion regime the deflected eruption begins at $t_1=184t_A$. At this time the apex of the flux rope is at $z\simeq 1.75$. 
The configuration of the system around the moment of the eruption is shown in Figure~\ref{Current-test} (right panels). The semitransparent green plane indicates the height $z=1.75$ where the decay index at the apex of the magnetic flux rope is $n\simeq 1.4$ (see Figure~\ref{Current-test}). If the photospheric motions are stopped at $t^{\ast} =180 t_A$, no eruption occurs.  However, if we stop the flows four Alv\'{e}n times later, a deflected eruption occurs. At time $t^{\ast}=184 t_A$, the axis of the flux rope has just reached the height where $n_{crit} \simeq1.4$, suggesting that this is the critical decay index in the `low' diffusion regime. Therefore, the critical values of the decay index presented in the previous section should be considered as an upper limit, with a margin of about 7\%.   

\begin{figure}
\begin{center}
\subfigure{\label{alpha1}
\includegraphics[width=.48\textwidth]{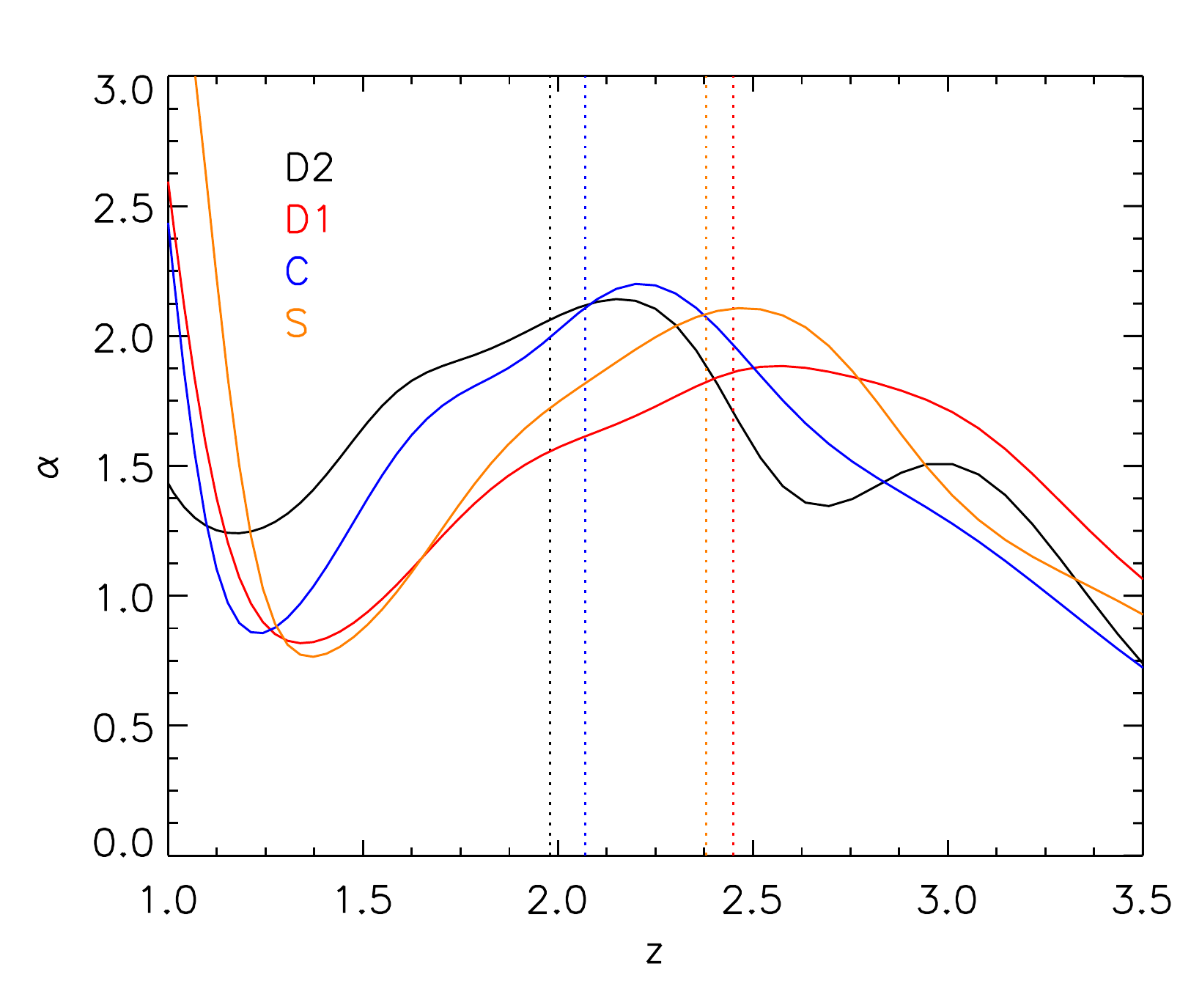}}
\subfigure{\label{alpha2}
\includegraphics[width=.48\textwidth]{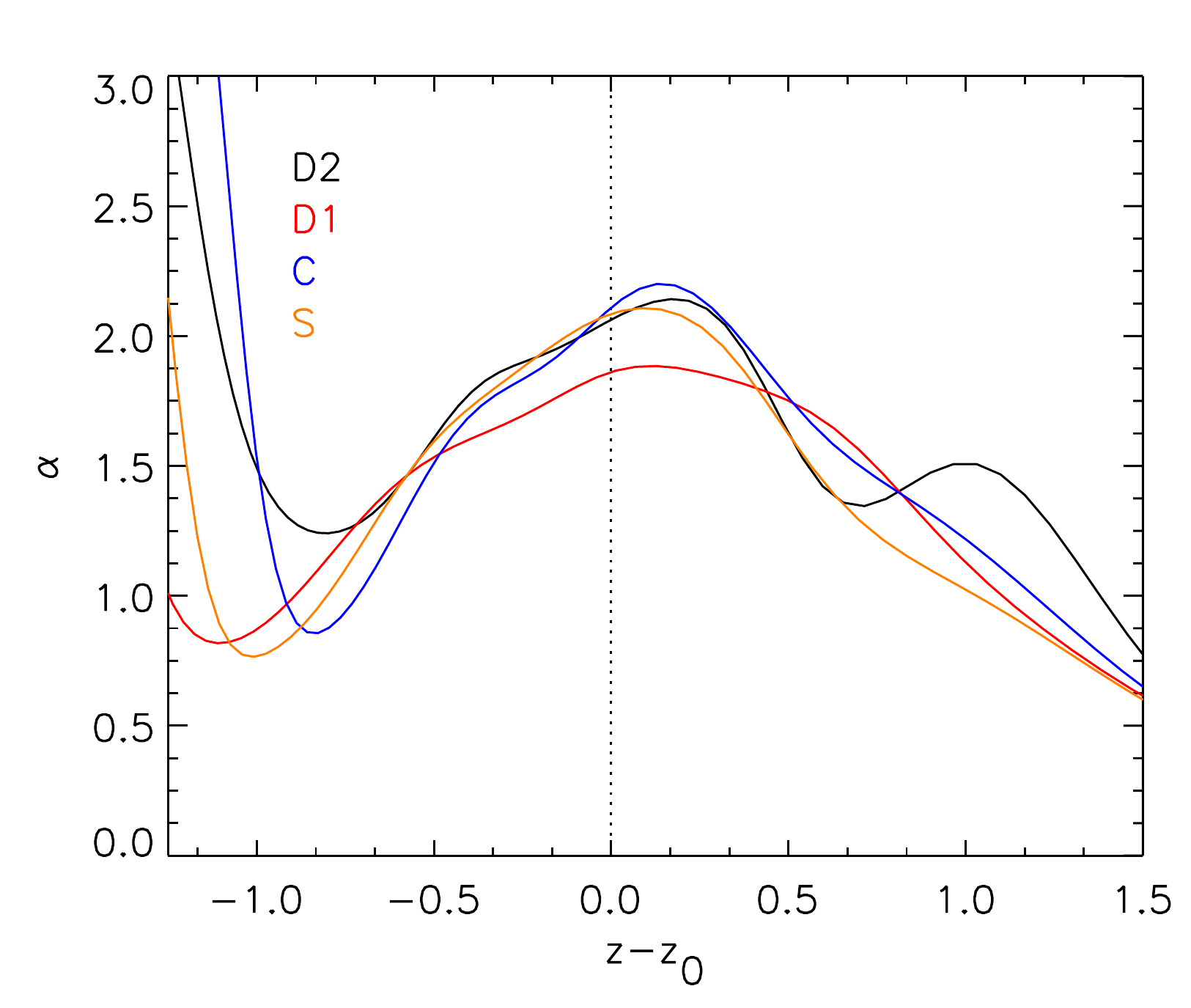}}
\caption{ \textit{Top panel:} Vertical dependence of  $\alpha= \mathbf{J} \cdot \mathbf{B} / B^2$ along a line passing through  the magnetic flux rope's axis. The vertical dotted lines indicate the height of the magnetic flux rope's apex for each simulation (color code) at the moment of the onset of the instability.  \textit{Bottom panel:} Same as top panel, but now the plot are shifted along the $z$ axis by an amount $z_0$, equal to the height of the axis of the magnetic flux ropes.}
\label{alpha}
\end{center}
\end{figure}

\section{Discussion and conclusion} \label{Conclusion}

We perform a series of numerical MHD simulations in order to investigate
the initiation and early evolution of flux-rope eruptions. We examine
how the initiation process depends on the photospheric magnetic
field distribution and the current profile of the
flux rope. In particular, we examine if the trigger of the eruption
is independent of the boundary flows that drive the system and form
the flux rope.

We perform four calculations using different photospheric driving
flows. Each calculation is split into three phases characterized
by the nature of the applied photospheric driver. In the first
two phases, driving flows are imposed in order to form a flux rope.
In the final phase, no driver is imposed, and the system is allowed
to evolve on its own.

In the first phase, we impose vortex motions (the same for all the
calculations)
around the iso-contours of the normal component
of the magnetic field. This flow profile generates electric currents in the
volume and causes the system to evolve into a quasi-force-free state.
At the end of this phase, the magnetic field
consists of a strongly sheared arcade near the PIL enclosed by an overlying
quasi-potential field. The flow profile in this
phase is chosen to build up shear in the system not to mimic
flows observed on the Sun. Although vortex motions are quite often
observed around sunspots, they are rarely so uniform and long lasting
(in terms of Alfv\'{e}n times) as the ones that we impose.

In the second phase, we impose four different classes of photospheric
motions.
Unlikely the previous phase, each of the four calculations is subjected to
a different flow profile. The profiles
are chosen to mimic different processes commonly observed during the
evolution of active regions, such as the deformation and spreading of
photospheric
magnetic field polarities. Similarly to what is observed on the Sun,
these motions have a
converging component, i.e., they tend to draw flux towards the PIL.
The flows are defined in terms of the magnetic
field gradient (see Section \ref{boundary} for details).
Since the magnetic field at the boundary is evolved with
the induction equation using the line-tied boundary conditions, the flows
are nonlinear both in time and space. Finally, all the imposed motions
are always sub-Alfv\'{e}nic (peaking at 5\% of the mean coronal
Aflv\'{e}n speed)
and the corona responds quasi-statically.

The flows at the photosphere draw flux towards the PIL, where
flux cancellation occurs because the magnetic diffusivity in
our model is finite. As a consequence, the topology of the magnetic
field changes
from that of a sheared arcade to that of a configuration with a
bald-patch separatrix.
During this phase, the applied photospheric flows transport magnetic
flux towards a three-dimensional current sheet, which results in the
formation of a flux rope through magnetic reconnection at the bald-patch.
As this process continues, the flux rope slowly rises due to the build
up of magnetic pressure. However, after a certain point, the system
undergoes a transition
to a dynamical regime characterized by the exponential growth of the
kinetic energy, and the flux rope erupts.

The third phase begins when the rope becomes unstable, at which
point we stop all photospheric driving. The instability point
is a priori unknown and we determine it by a series of relaxation runs
(see Section \ref{relax}).

In summary, for a complete calculation, we drive the system to
form a flux rope, which we then bring to the threshold of
instability. At this point we stop driving the system and let it
evolve freely to an eruption.

We study the evolution of the system in the framework of the torus
instability scenario,
because it gives a well defined threshold that can also be applied to
observations. For
quasi-statically evolving magnetic fields, it has been shown that
the catastrophe and torus instability scenarios give the same prediction
for the onset of the loss of equilibrium or the torus instability
\citep{Dem2010, Kliem2014}.
The same is true when the torus instability criterion is compared with
analysis based on the energy of the semi-open field associated with a
given magnetic field distribution at the boundary \citep{Ama2014}.

The height of the flux rope when it erupts depends
on how the photospheric magnetic field is evolved.
Two interesting examples are `Run~D1' and `Run~D2'. For these two
calculations the apex of the magnetic flux rope at the onset of the
eruption is located at $z\simeq2.45$ and $z\simeq1.95$ respectively.
\cite{Tor2007}
have shown that the height of the onset of the eruption increases when
the distance between the photospheric charges increases. As can be deduced
from Figure~\ref{v-field}, the central part of the flux distribution for
`Run~D1' is essentially unmodified, resulting in a highly concentrated
flux distribution in the central part of the polarity. However, for
`Run~D2', the magnetic flux is more uniformly distributed within the
whole polarity. As a result, the baricentrum of the magnetic flux is
closer to the PIL for `Run~D2' than for `Run~D1', which results
in a dipole with a smaller length scale and, therefore, a lower
critical height for the onset of the torus instability.

Figure~\ref{alpha} (top) shows the profile of $\alpha= \mathbf{J} \cdot
\mathbf{B} / B^2$
along a vertical line passing through the apex of the axis of the magnetic flux
rope at the moment of the onset of the instability. Two different
critical heights can be seen in the figure. `Run~C' and `Run~D2' both
have a critical
height of $z\simeq2$, while `Run~S' and `Run~D1' both have a critical
height of $z\simeq2.4$.
Figure~\ref{alpha} (bottom) shows the same $\alpha$-profiles, but shifted
by the height of the axis of the magnetic flux rope. This allows a direct
comparison of the different curves. All the flux ropes are quite similar,
but `Run~D1' displays some minor differences. A comparison between
`Run~C', `Run~S', and `Run~D2' shows that similar flux ropes have
different critical
heights. In addition, `Run~D1' and `Run~S' have similar critical heights
($z\simeq2.4$), but their $\alpha$ profiles are different (within
the limitations of a 1D plot). We find that what is crucial for the
onset of the eruption is not
the actual height of the flux rope, but the fact that the flux rope
approaches the regions
where the decay index is close to the predicted critical range
$n_{crit}=1.4-1.5$.
This provides strong evidence in favor of the torus instability scenario.

In a current-wire treatment of the torus instability, the morphology
of the current channel influences the critical value of the decay index.
For example, the critical decay index of a straight wire is typically
smaller than a circular wire. For our simulations, we find that the
only case with a noteworthy difference in the critical decay index is
`Run~D2', but its flux rope’s current profile is not significantly different from
`Run~C,' and `Run~S'. We stress that our model flux ropes are quite similar (see
Figure~\ref{3DView}), however, it is possible that a geometrical effect may have also
played a role in the eruption, although it cannot be easily disentangled in our
asymmetric configuration. Nevertheless, it is interesting that `Run~D2' is the case
that displays a lower value of the critical decay index and is also the one
where the convergence flows induce the most significant modification of
$B_z$
at the boundary. The lower values of the critical decay index in this
latter simulation may be due to the different evolution of the bald-patch
separatrix, that possibly induces different line-tying effects.

Figure~\ref{3DView} shows a local enhancement in the current density in
the whole volume of the flux rope. This suggests that if one wanted to
compare our flux rope model with the wire models, one should consider an
almost circular, relatively thick current channel. In this context,
\cite{Dem2010} predicted a critical value of the decay index that depends on
whether or not the current channel expands during the perturbation. The
value that we find, $n_{\text{crit}}\simeq1.4-1.5$, would be compatible
with a flux rope that has constant current during the perturbation.
However,
the simulation without the `increased' diffusion showed that this range
could be up to about 7\% lower, that is, $n_{\text{crit}}\simeq1.3-1.4$.
This range would also be compatible with circular current wires with a
time-varying current. Various processes can influence the
evolution of the current within the flux rope as it evolves.
During its slow quasi-static rise, the flux rope expands, which causes
its twist per unit length to decrease. As a result, the volume current
in the rope also decreases. However, during the formation process, and
subsequently the eruption itself, magnetic reconnection at the bald-patch,
and subsequently at the flare current sheet, clearly injects some sheared,
current-carrying, magnetic flux into the flux rope, and this must
eventually
increase the volume current therein. If the two opposite effects
balance each other, then the volume current of the flux rope may not
vary too much, which will result in an instability threshold similar to the
one for currents rings with a modest temporal variation of the current,
that is, $n_{crit} \in [1.2-1.5]$ \citep[][]{Dem2010}.

The analysis presented in this paper suggests that the trigger of the
eruptions is the
torus instability regardless of the exact morphology of the magnetic
flux rope and
of the photospheric evolution of the active region. However, several
phenomena such
as line-tying, magnetic reconnection at bald-patches and different
evolution of the
photospheric magnetic field affect the critical decay index for the
onset of the instability,
resulting in a `critical range' rather than a `critical value'. We
speculate that the decrease
of the current due to the flux rope expansion and its increase due to
the bald-patch
reconnection compensate, resulting in a critical range that is not too
different from the
analytical predictions of the wire models.

\acknowledgements

The work of F.P.Z. is funded by a contract from the AXA Research Fund. F.P.Z. is a Fonds Wetenschappelijk Onderzoek (FWO) research fellow on leave. S.A.G. acknowledges the financial support of the DIM ACAV and R\'{e}gion Ile de France. This work was granted access to the HPC resources of MesoPSL financed by the R\'{e}gion Ile de France and the project Equip@Meso (reference ANR-10-EQPX-29-01) of the programme Investissements d' Avenir supervised by the Agence Nationale pour la Recherche.

\section{Appendix A}

\begin{figure*} [!h]
\begin{center}
\subfigure{\label{RunC-energy}
\includegraphics[width=.55\textwidth,viewport= -10 0 550 390]{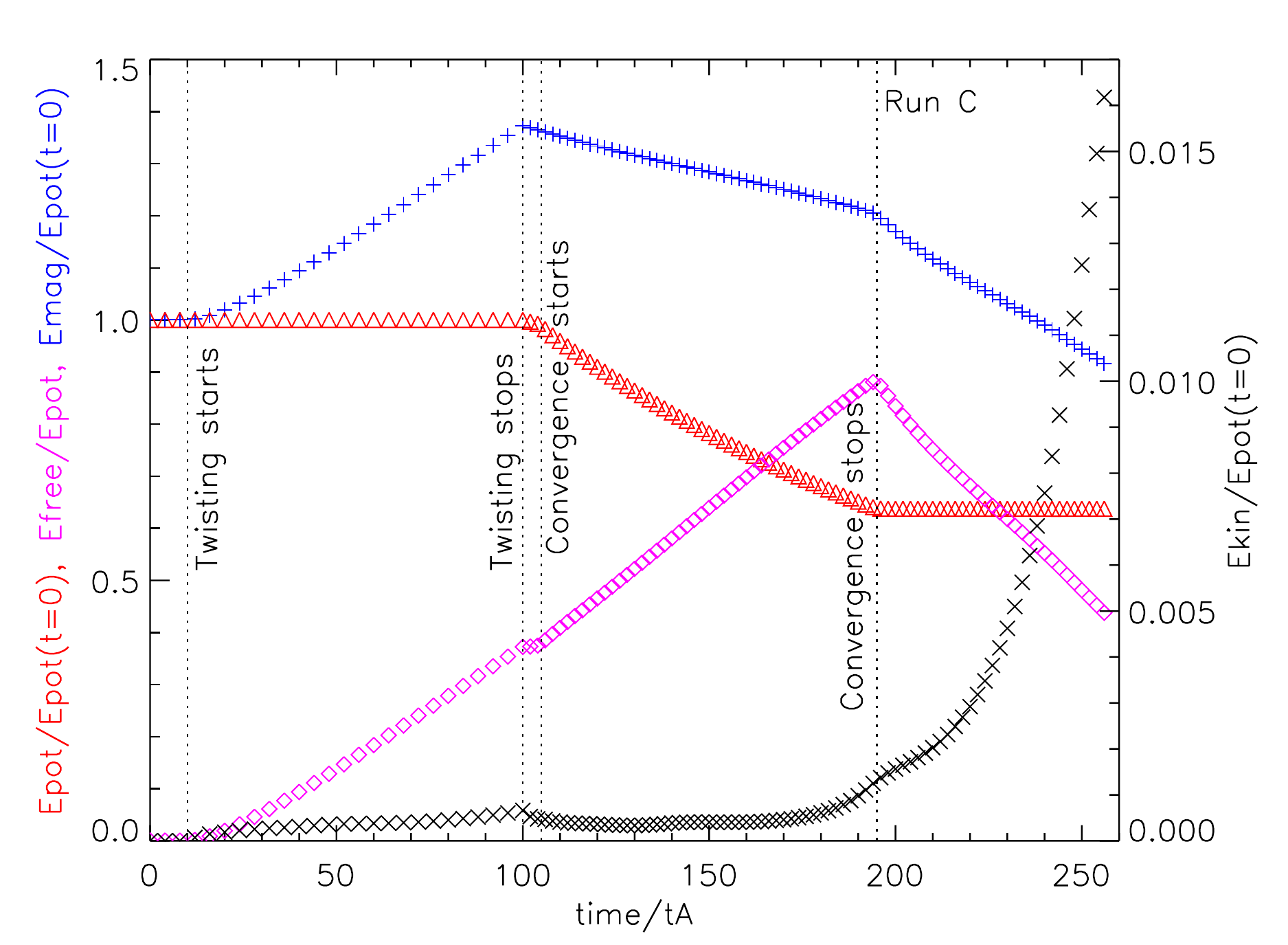}}
\subfigure{\label{RunS-energy}
\includegraphics[width=.55\textwidth,viewport= -10 0 550 390]{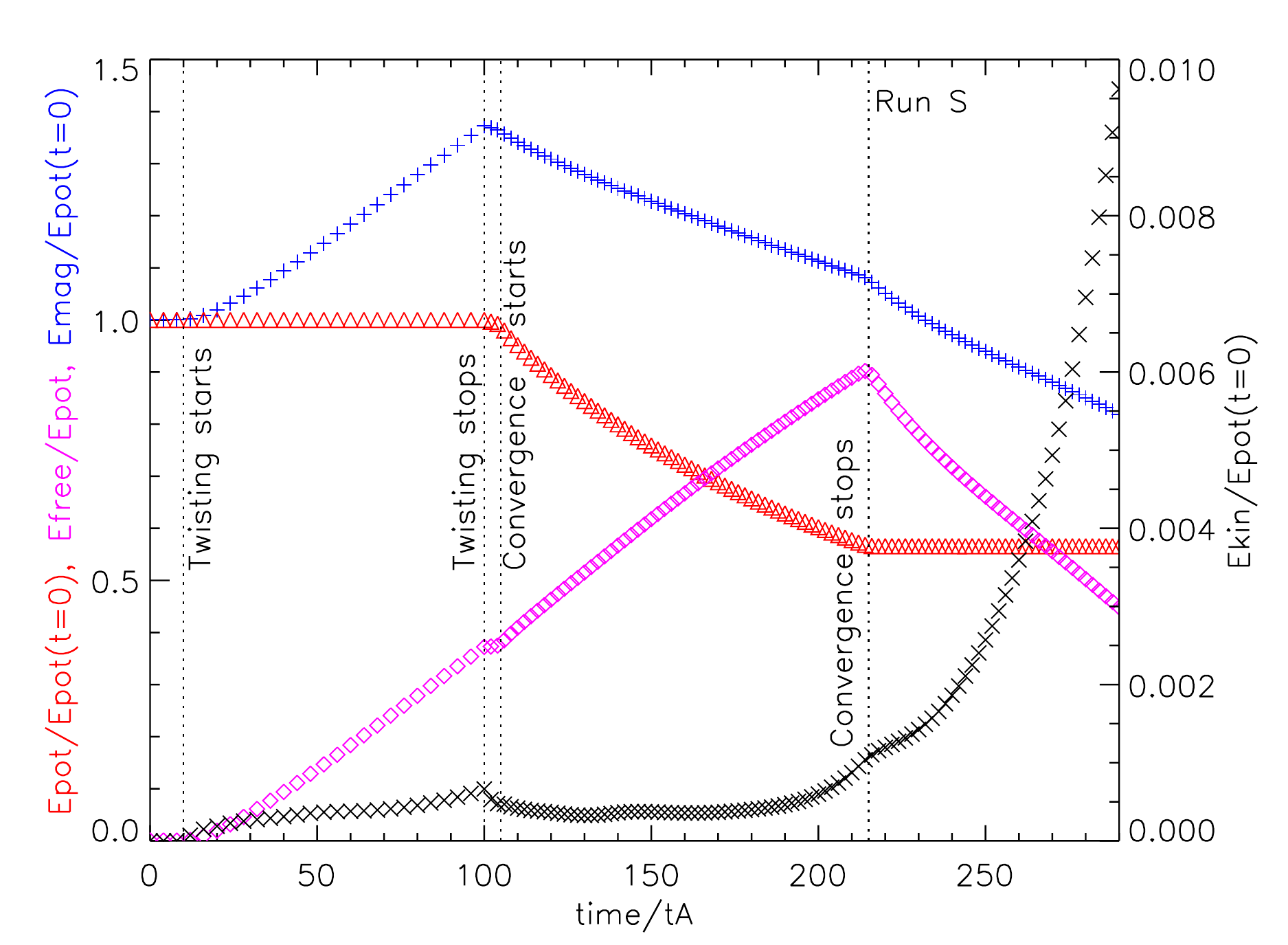}}
\subfigure{\label{RunD2-energy}
\includegraphics[width=.55\textwidth,viewport= -10 0 550 400]{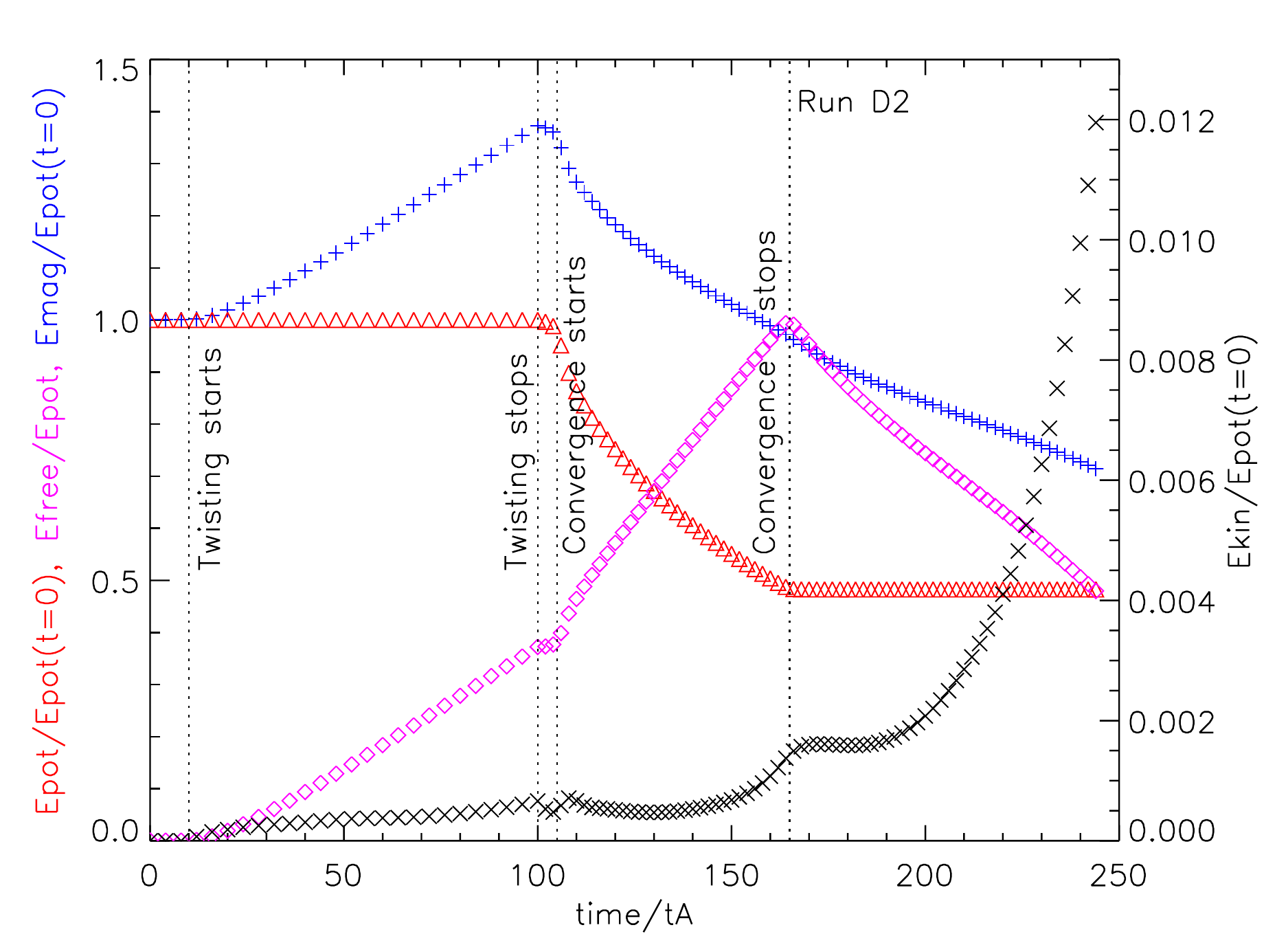}}
\caption{Temporal evolution of the free magnetic energy in the system (purple `$\diamond$') as well as of the kinetic (black `$\times$'), potential (red `$\bigtriangleup$') and total (blue `$+$') magnetic energy of the system normalized to the energy of the initial (potential) magnetic field for `Run~C', `Run~S' and `Run D2' (from top-left to bottom-right, respectively). The magnetic energy of the initial potential field is $E_{\text{pot}}(t=0)=174.4$.  }
\label{Energy}
\end{center}
\end{figure*}


%

\end{document}